# A general relationship between extinction risk and carrying capacity

**Authors**

Thomas S. Ball[1], Ben Balmford[2], Andrew Balmford[1], Daniele Rinaldo[2], Piero Visconti[3], Barry W. Brook[4], and Rhys E. Green[1].

**Affiliations**

1. Conservation Research Institute, Department of Zoology, University of Cambridge, Cambridge, UK.
2. Land, Environment, Economics and Policy Institute, Economics Department, University of Exeter Business School, Exeter, UK.
3. Biodiversity, Ecology and Conservation Research Group, International Institute for Applied Systems Analysis (IIASA), Vienna, Austria.
4. School of Natural Sciences, University of Tasmania, TAS 7005, Australia.

**Abstract**

Understanding the relationship between a population's probability of extinction and its carrying capacity frames conservation status assessments and guides efforts to understand and mitigate the ongoing biodiversity crisis. Despite this, our understanding of the mathematical form of this relationship remains limited. We conducted ~5 billion population viability assessments that jointly converge on a modified Gompertz curve. This pattern is consistent across >1700 distinct model populations, representing different breeding systems and widely varying rates of population growth, levels of environmental stochasticity, adult survival rate, age at first breeding, and initial population size. Analytical treatment of the underlying dynamics shows that few assumptions suffice to show that the relationship holds for any extant population subject to density-dependent growth. Finally, we discuss the implications of these results and consider the practical use of our findings by conservationists.

**Introduction**

How does the extinction risk of a population shift with environmental change? We propose a framework with which to consider such changes, focusing on the impact of reductions in carrying capacity on extinction risk. This question underpins several core conservation problems. Which species are at greatest risk and therefore need most urgent attention? Which sites, if protected, offer the best prospects of retaining populations of concern? And how are changes in area of habitat or climate envelopes likely to affect extinction risk? Despite substantial progress, we still lack a general model for the shape of the relationship between a population's probability of extinction and the carrying capacity of the environment in which it lives. Through extensive simulations we uncover a generally applicable model relating extinction risk to carrying capacity and show that it has a theoretical basis. We hope that our discovery might be used to inform future conservation efforts, and support future efforts to quantify and mitigate anthropogenic harms to life on Earth.

The applied importance of assessing extinction risk emerged from the foundational work of Mace and Lande (1991), who developed quantitative, repeatable procedures now applied to assess the conservation status of >160000 species (www.iucnredlist.org). In parallel with this, researchers have modelled expected time to extinction extensively (Ovaskainen 2010; Melbourne 2008; Iwasa 2000; Hanski 2000; Lande 1993), and the minimum size needed to reduce the risk of extinction below an acceptable threshold within a specified period. Here, 'population' is used here to describe a set of individuals interacting with the same environment; a species' global meta-population might comprise several such populations. The small population paradigm proposes that the probability of extinction ($P_E$) rises sharply as carrying capacity ($K$) decreases due to reinforcing feedbacks (Caughley 1994), yet surprisingly little attention has been given to the precise form of this relationship. Several studies (Thomas et al. 2004; Phalan et al. 2011; Strassburg et al. 2012, 2018; Armsworth et al. 2020) have assumed, by analogy with the species-area relationship, that the probability of survival $P_S$ (i.e. 1 - $P_E$) increases with $K$ according to a power law. Other studies suggest a sigmoidal relationship (Dushoff 2000; Dennis 1991). Brook et al. (2006) and Hilbers et al. (2016) used simulation models to explore the effects of changes in $K$ on $P_E$. Building upon this, Wolff et al. (2023) re-analysed Hilbers et al. (2016) data to examine the $P_E$ - $K$ functional form. They find that a Gompertz curve fits most mammals, although they provide no assessment of whether the Gompertz is the true underlying relationship, nor how the shape changes with different parameter and demography assumptions.

We argue that it is essential to understand the $P_E$ vs. $K$ relationship and its generality, because it will strongly influence the effects on biodiversity of recent and ongoing habitat destruction and degradation. Here, we tackle this problem using simulation models characterising a range of population processes across varying demographic parameters. Remarkably, we discover that one particular curve – a modified Gompertz curve – provides an exceptionally good fit to our simulation results across a wide range of model parameters and structures. We go on to provide a theoretical explanation for why the curve takes this form, and, with the assumption of density-dependent growth, show that it applies to all extant populations. We believe these discoveries, and the underpinning framework, substantially enhance our understanding of, and thus capacity to mitigate, the dynamics of the unfolding extinction crisis.

**The shape of the curve**

We conducted simulations using four population models with increasingly complex structures, each allowed to run for 100 years, a timeframe chosen for its congruence with the IUCN Vulnerable category (the least threatened at-risk category; IUCN 2001). All our models assumed density-dependent growth between years, reflected by a logistic relationship between population growth rate and current population size relative to carrying capacity. Model A treated males and females as separate sub-populations with independent growth rates, but a shared environmental realisation. In Model B the growth rate of males and females was dependent on the total number of both sexes combined. Model C built on Model B by incorporating biparental care, so unpaired adults do not reproduce. This model also required assumptions regarding adult annual survival rate and age of first breeding. For Models A, B, and C, annual environmental stochasticity was temporally independent. Model D added temporal autocorrelation in environmental stochasticity between years, making consecutive good or bad growth rates years more likely. Models A and B used two input parameters: the maximum population growth rate ($r_{\max}$) and extent of environmental stochasticity ($\sigma$). Model C had two additional inputs: $S_a$ - the annual survival probability for an adult in the population, and $B$ - the age at first breeding. Model D had a further input parameter $Z$, which controlled the degree of temporal autocorrelation. Detailed descriptions of the four models, their parameters, and simulation setup are given in Supplementary Appendix A.

We simulated all four models across a wide parameter range, for values of carrying capacity *K* ranging from 1 to 3 million (for an explanation of input parameter ranges see Supplementary Appendix B). We contend that, given that species' lifetimes are typically in the range of $10^5$-$10^7$ years (Barnosky et al. 2011), currently extant species are so because the dynamics of their populations are sufficient to ensure their medium-term persistence (in the absence of external pressures). We therefore only consider sets of input parameters as 'viable' if $P_E \approx 0$ in 100 years at our maximum simulated *K*.

We initialised 10,000 populations for each parameter combination and value of *K*, calculating the observed probability of survival ($P_S$) as the proportion of replicates still extant at 100 years. We then investigated the relationship between the probability of survival $P_S$ and the carrying capacity $K_s$ to test whether a common functional relationship exists between the two. We did so by modelling $P_S$ as a function of $K$ (denoted $P_S(K)$) using a Gompertz curve, modified through the addition of an additional shape parameter ($\gamma$), which primarily acts as a skew factor. Larger $\gamma$ steepens the slope and sharpens the inflection. Extinction probability $P_E$ as a function of $K$ (denoted $P_E(K)$) within 100 years is then given by $1 - P_S(K)$; this relationship is thus a transformed and modified Gompertz curve (hereafter "modified Gompertz"), which is an asymmetric sigmoid. Formally:

$$P_S(K) = \exp(-\exp(a + b\, K^\gamma)),$$
$$P_E(K) = 1 - P_S(K)$$

(1)

where $a$, $b$, and $\gamma$ are constants. Fitting this model to simulations for each viable parameter set and population model (A-D), using a process described in Supplementary Appendix C, generated $r^2$-values every one of which exceeded 0.995. Figure 1 shows an example curve for each of our four types of population and illustrates the extremely close fit of the modified Gompertz despite marked differences in model assumptions. This comparison also makes the noteworthy point that moving from models A-D – that is to say adding realism - increases the probability of extinction (i.e. the $P_E(K)$ curve shifts to the right).

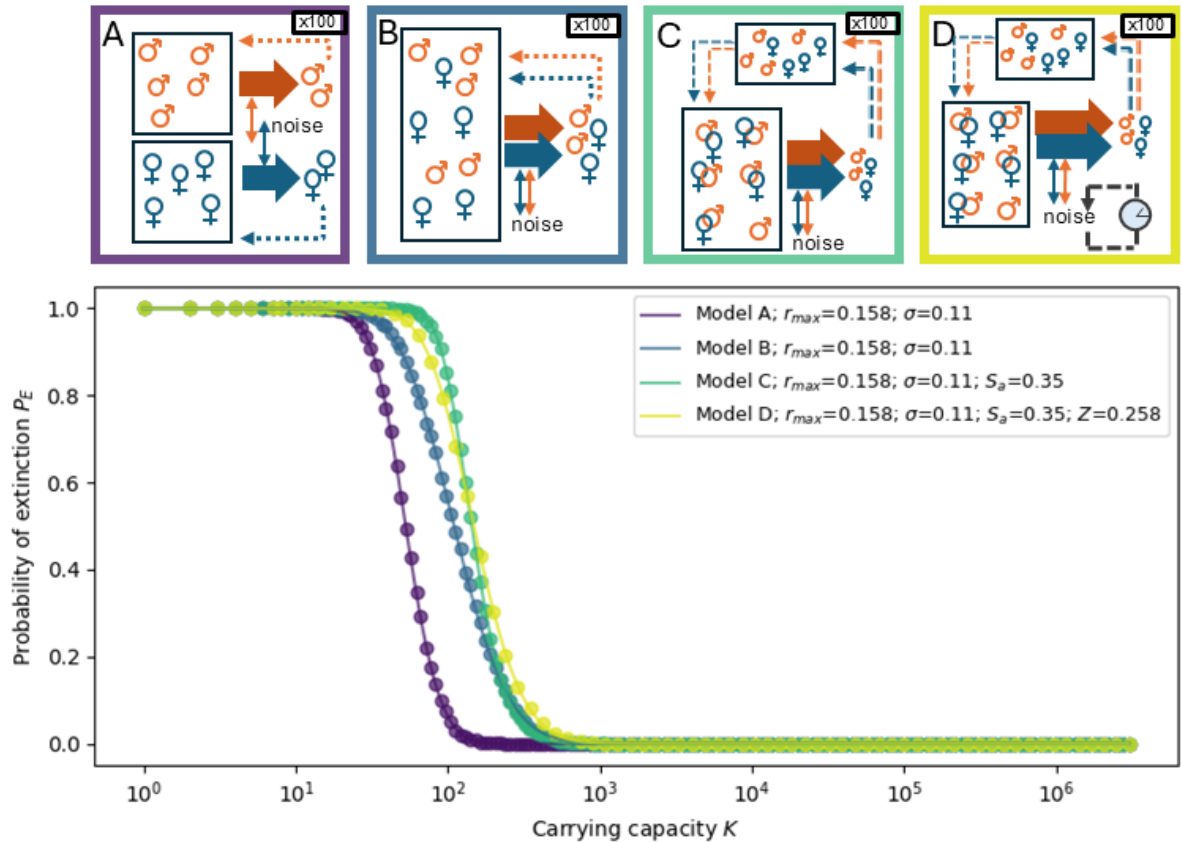

Figure 1. Modelled probabilities of extinction (circles) for a range of carrying capacities and fitted modified Gompertz curves (lines) for Models A, B, C, and D, along with a diagrammatic representation of each model. Here, $r^2 > 0.9999$ for all four curves. In each diagram, the individuals contained within the sub-box(es) represent the 'current' population, with those outside and the various arrows representing the mechanisms by which at the population size changes for the next time step. The large arrows represent the growth rate at a given time which is subject to environmental stochasticity as shown by the 'noise' arrows. Model A with male and female growth rates (large arrows) dependent only on the number of individuals of each sex respectively. Model B with growth rate for each sex being dependent on the number of individuals in each sex together. Model C with a 'delay' to growth caused by the time taken for juveniles to reach breeding maturity, and Model D: the same as Model C, but with temporally autocorrelated environmental stochastic noise. We used mid-range values for the input parameters: $r_{\max}$=0.158, $\sigma$=0.11 for all four models, $S_a$=0.35 for Models C and D, and $Z$=0.258 for D.

**The modified Gompertz is generally applicable**

A modified Gompertz curve thus generally describes the relationship between probability of extinction and carrying capacity across a broad range of input parameters and model types. To further understand this relationship, we considered cases where population size is initialised at a value other than carrying capacity (Figure 2). This is of biological relevance, for example, when an extirpated population is re-introduced into its former range (so the initial population is much smaller than carrying capacity), or an area of a species' habitat is suddenly lost (so that the initial population exceeds carrying capacity). We find that under these circumstances the relationships continue to follow modified Gompertz curves – albeit with slightly different shapes. Using Model

A, but with initial population size ($N_0$) set at a fraction of $K$ (Figure 2a), larger carrying capacities are required to prevent increases in the extinction risk when the initial population is a smaller proportion of $K$. Figure 2b shows how the probability of extinction for a fixed initial population size responds to changes in carrying capacity. Moving from right to left along each line effectively charts how reductions in carrying capacity would impact upon an existing population's extinction risk. Note that, owing to negative density-dependence, larger population sizes have higher probabilities of extinction for the same $K$, implying that otherwise stable populations experiencing a sudden reduction in their carrying capacity ($N_0 > K$) are at a heightened risk of extinction until the population stabilises at the lower $K$.

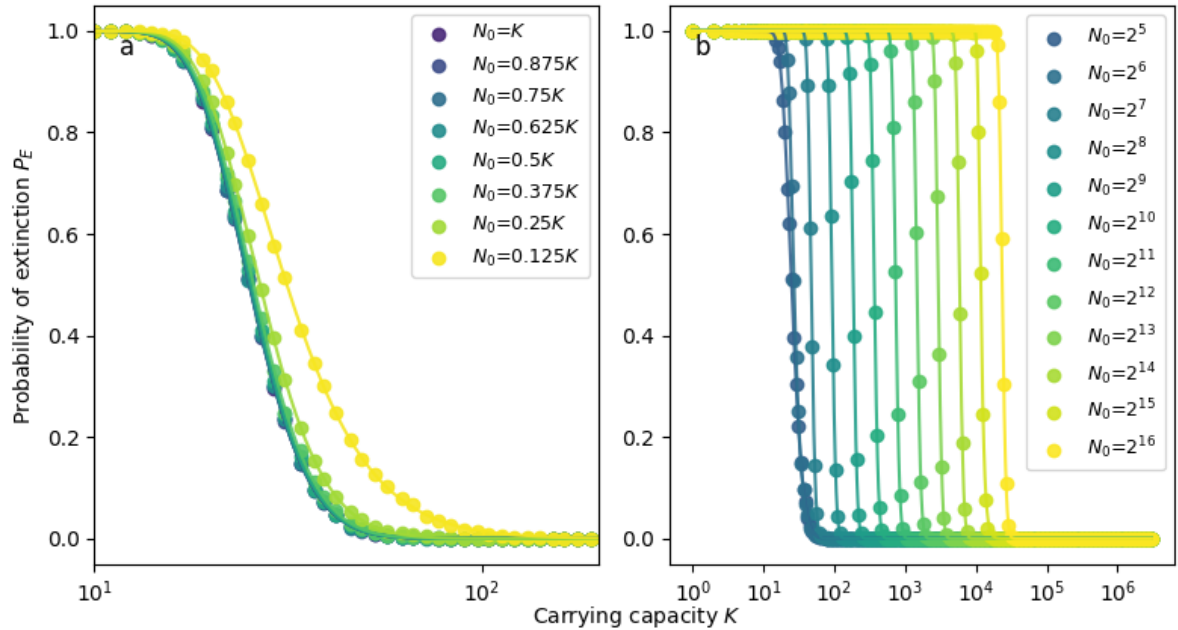

Figure 2. Empirical probabilities of extinction (circles) for a range of carrying capacities and fitted modified Gompertz curves (lines) when initial population size ($N_0$) in Model A ($r_{max}$=0.56, $\sigma$=0.15) differs from the carrying capacity $K$ either (a) as some proportion of $K$ or (b) is an absolute number independent of $K$. Note that in (a), all changes in $P_E$ occur between 10 and approximately 100, and that in (b) $P_E$ begins to drop when $K$ is approximately the same order of magnitude as $N_0$. Associated $r^2$ values for the modified Gompertz curves remain extremely high ($r^2$>0.9999 for all curves). In (b), we conducted simulations with $N_0$ < 32, but they were discounted as they did not meet our validity criteria (i.e. $P_E$ = 0).

We next tested the modified Gompertz model's performance when populations were subject to Allee dynamics, with modified growth rates at low population sizes. 'Extinction vortices' (Gilpin and Soulé 1986), mate limitation, or cooperative defence or feeding interactions (Dennis 1989), are examples of mechanisms for negative and positive density dependent growth that have been observed to affect the growth rate of small populations. These mechanisms produce classic Allee dynamics (Dennis 2002), in which an unstable population equilibrium exists such that the density dependence of the growth rate is positive above, and negative below, a critical population size threshold. We modified Model A to include Allee effects (see Supplementary Appendix D for details), then conducted simulations varying both the critical threshold population $\theta$, where $\theta$ < $K$, and the strength parameter of the effect $\upsilon$, where $\upsilon$ > 0. We found that the presence of Allee effects resulted in a quasi-extinction threshold at $\theta$ (as opposed to 1), below which it is much harder (though not impossible) for the population to escape extinction. This shifted curves to the right, extending the at-risk section of the curve into higher carrying capacities, alongside steepening the point of inflection. We found that the modified Gompertz function fitted the results to the same high degree of accuracy as in the main set of results ($r^2 > 0.9998$, see Supplementary Figures D1 and D2).

**Applying the modified Gompertz to empirical data**

Real-world data on probability of extinction as it relates to carrying capacity are scarce. However, we were able to test the modified Gompertz curve on three empirical case studies to confirm its applicability to real populations. The first was an observational study of isolated populations of bighorn sheep (*Ovis canadensis*) in the Western USA (Berger 1990). We found that the modified Gompertz was able to describe the data with $r^2 > 0.88$ across each set of results (see Supplementary Appendix E for detailed methods and results). The second was a modelling study on the persistence of New Zealand brown mudfish (*Neochanna apoda*). White et al. (2017) used proprietary population viability analysis software to assess the viability of predicted local populations in the face of highly variable environmental stochasticity. Fitting the modified Gompertz curve to their data yielded very good fits ($r^2 > 0.98$). Our third test case was a population modelling study parameterised using population-specific demographic rates estimated from observations from a 12-year study of brown bears (*Ursus arctos*) in Yellowstone National Park. Shaffer and Samson (1985) reported estimates of probability of extinction within 100 years for seven hypothetical carrying capacities. The modified Gompertz fit was again excellent ($r^2 = 0.9991$).

**The theoretical basis for the modified Gompertz**

To better understand the apparent generality of the modified Gompertz curve, we complemented our simulations with a theoretical analysis of Model A, whose relative simplicity allows for analytical tractability (all technical details and formal derivations are set out in Supplementary Appendix F). We focused on the continuous-time, logistic stochastic differential equation (SDE) and studied its properties. We then obtained its expected time to extinction as a function of $K$ and defined the parameter space that allows extant species to exist. In accordance with our simulations, we focused on a population starting at carrying capacity and showed how its probability of survival $P_S$ can be well approximated by a Gompertz curve in $K$. We achieved this result by approximating the transition density of the logistic SDE with that of an Ornstein-Uhlenbeck process centred on carrying capacity, then integrating it over all achievable (i.e. non-extinct) states.

The intuition behind this result is that for all viable parameter sets, the population tends to revert toward carrying capacity after stochastic fluctuations. As carrying capacity increases, the role played by the initial fluctuation becomes increasingly less relevant and the probability of long-run persistence increases. Indeed, this theoretical analysis reveals the Gompertz relationship to be general, to Models B to D, but more importantly to all populations - contingent on just two assumptions. First, that population growth rate is governed by negative density dependence, such that the population cannot grow infinitely. This negative density dependence could take any form and is not restricted to logistic growth. Second, that for large enough carrying capacity the population demographics give rise to a steady state population above 0 (i.e. its demographic parameters are contained in the viable parameter set we suggest describes extant species). Even the inclusion of Allee effects, which push populations below a threshold size toward extinction, does not change the Gompertz nature of the curve. Given these assumptions, the relationship between extinction risk and carrying capacity is well approximated by a standard Gompertz curve: a version of Equation (1) for which the shape parameter $\gamma = 1$.

However, while fitting this standard Gompertz to our simulation results yields high $r^2$ values (>0.994, >0.995, >0.992, and >0.971 for models A, B, C, and D respectively), the curve appears to systematically deviate from our simulated results (supplementary Figure G1), possibly arising from the approximations made to arrive at the standard Gompertz. The size of this deviation increases as the inflection point of the curve moves into higher $K$. By modifying the Gompertz curve with the shape parameter $\gamma$ and fitting the modified curve to the same simulated data we achieved much closer fits to the simulated data - obtaining higher $r^2$ values for all simulated curves (Supplementary Figures G2-5), as well as for the empirical datasets (Supplementary Section E). We propose that the modification achieves better $r^2$ values not by virtue of higher dimensionality but systematically. If the improved fit was a result of the former, one would expect the values for the additional parameter to be distributed about 1. However, the value of the shape parameter $\gamma$ is consistently below 1 across all models, with median values of 0.85, 0.46, 0.34, and 0.32 for models A through D respectively (see Supplementary Figure G6).

**Prioritising conservation interventions**

As one of many implications of understanding the extinction risk- carrying capacity relationship, consider priority-setting among competing conservation alternatives. If the primary goal is to limit extinctions, which recent work has indicated to be the preference of both experts and a general audience (Meier et al. 2024), priority for interventions - other things being equal - could be given to those for which a marginal increase in carrying capacity is likely to have the largest reduction in extinction risk, i.e. those for which the first derivative of the curve is greatest: the point of inflection of the curve.

We note that the point of inflection on the modified Gompertz curve differs significantly from that of other curves that have been used to inform conservation. Previous studies have discovered that the relationship is sigmoidal in some way, though did not explore the functional form of the curve (Dushoff 2000; Dennis 1991). Figure 3 compares a best-fit logistic curve, perhaps the most obvious sigmoid to test, with the modified Gompertz. A logistic curve scaled between 0 and 1 has a point of inflection at $P_E = 0.5$ by construction, but our simulated results have inflection points consistently below 0.5. A power law function was derived indirectly by Lande (1993) and has been used in a range of studies (Thomas et al. 2004; Phalan et al. 2011; Strassburg et al. 2012, 2018; Armsworth et al. 2020; Eyres et al. 2025), but as Figure 3 shows, even a best-fit power law curve deviates strongly from our simulated results. The formulation of the power law dictates that the greatest rate of change and thus return on conservation effort will arise at when $P_E = 1$. In addition, for all but the most at-risk populations, the power law significantly underestimates the rate at which extinction risk changes with $K$ when compared with the modified Gompertz.

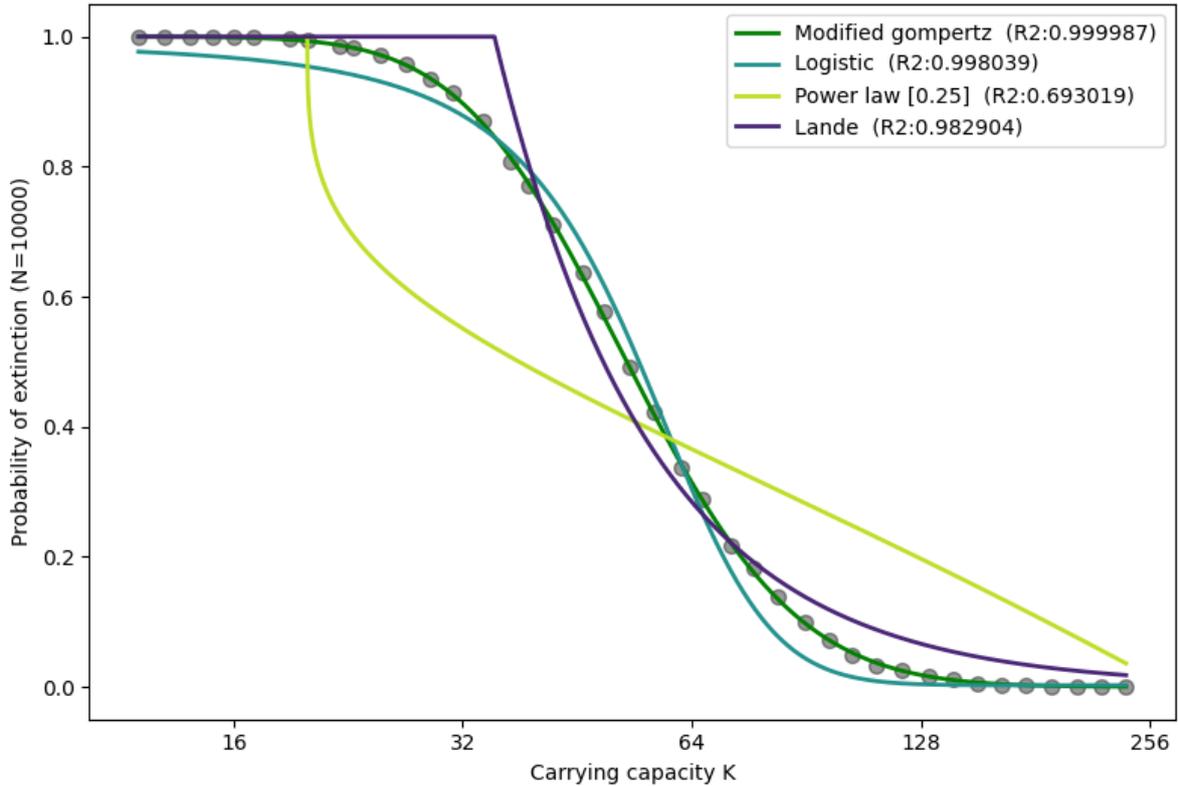

Figure 3. Comparison best fits for modified Gompertz, logistic, power-law, and Lande power-law (1993) curves when fitting to an example of Model A data ($r_{max}$=0.158, $\sigma$=0.11). The logistic is a standard logistic function with a maximum value of 1. The Lande curve is derived from Equation 8b of Lande (1993) and has the form $P_E \propto (1/K)^{|a|+a}$ where $a$ is a positive constant. The power-law curve here takes the form $1 - P_E \propto (K + X_o)^{0.25}$ where $X_0$ is a horizontal offset to align the curve with the first $P_E$ value not equal to 1.0, and was fitted by constraining the data to only points where 0 < $P_E$ < 1. The value 0.25 is a commonly used exponent (Eyres et al. 2025; Thomas et al. 2004).

For all our simulation models, the inflection point of the modified Gompertz occurs when the probability of extinction is greater than 50%, with median $P_E$ values at the point of inflection being 0.65, 0.75, 0.74, 0.75 for Models A, B, C, and D respectively (Supplementary Figure H1). These results, which are remarkably consistent across parameter combinations, suggest that conservation efforts that affect species' effective carrying capacity might yield the highest benefits for populations for which the calculated $P_E$ is between 0.65 and 0.85. An important caveat is that potential interventions will of course vary in other ways too – particularly in net costs, which may often be higher in situations where anthropogenic pressures are such that populations are closer to extinction. Thus, we contend not that highest overall conservation priority should be given to species with $P_E$ between 0.65 and 0.85, but that ecological triage thresholds should consider extending to such a range.

**Towards estimating curve shapes for real-world species**

In an effort to maximise the utility of our results, in a final set of analyses we explored how far a population's extinction curve can be estimated from broadly available life-history parameters, namely population growth rate and environmental stochasticity, which are input parameters to all our simulation models. We began by defining $K_{10}$ – the carrying capacity for which a population has a 10% probability of extinction within 100 years; this threshold is of particular conservation relevance as the IUCN Red List's Criterion E categorises species below it as Vulnerable (IUCN 2001; for discussion of how this and other thresholds manifest in our simulated populations see Supplementary Appendix I). We then built linear models to predict the two parameters of the standard Gompertz curve from our simulation input parameters (Supplementary Appendix J). We were unable to go beyond this to arrive at the full form of the modified Gompertz curve due to an over-identification problem. Nevertheless, this means that, given a reasonably accessible set of life history parameters for a species, we can both approximate the curve describing how its extinction risk changes with carrying capacity and predict carrying capacities associated with critical extinction thresholds. In a conservation context we suggest this approach might be usefully applied, in the first instance, to rapid species assessment, categorisation, or category validation.

**Discussion**

Using simple population dynamic models, we have discovered that the relationship between population carrying capacity and probability of extinction is extremely well approximated by a modified Gompertz function. This finding is robust across billions of simulations of many different permutations of the parameter inputs and modelling assumptions and appears to fit well to the findings of the small number of empirical studies to which we could apply it. We have shown theoretically why the curve is approximated by a Gompertz curve, and crucially that this is general across all extant species as long as they face some form of density dependence.

For the great majority of our model populations the probability of extinction decreases to near-zero very quickly as $K$ increases. Many real-world species often have global population sizes in the millions (Callaghan 2021). Some species may be well-represented by a single population of the type we have simulated here, whilst dispersal, migration, habitat and resource fragmentation, amongst other things, will necessitate a more complex approach for others. Further, our models are of individual populations and may not apply in the same way at global scales. Successful modelling of global metapopulations will require understanding of the dynamics of subpopulations as well as their interactions; we believe that we have gone some way to addressing the former in this study.

This paper follows the small population paradigm in focusing on the role of stochasticity in driving populations extinct. Future research could extend our modelling framework to assess the impacts of human-driven deterministic changes – such as the increases in extinction risk caused by anthropogenic reductions in population growth rates or adult survival, or ongoing declines in habitat available and corresponding carrying capacity, as well as increases in environmental stochasticity. Understanding the likely extinction impacts of such changes – acting either independently or synergistically – is crucial if we are to begin to halt and reverse biodiversity loss. To this end, we believe this study represents a significant step forward in understanding the dynamics of the extinction crisis and in prioritising, guiding and evaluating efforts to mitigate it.

**Data availability**

All generated data described in this manuscript is available online (DOI: 10.5281/zenodo.14419062). Data on generation lengths for bird species used to inform the choice of simulation parameters as described in Supplementary Methods C is available in the supplementary materials of Bird et al (2020), DOI: https://doi.org/10.1111/cobi.13486.

**Code availability**

All code for running simulations, conducting analysis, and producing figures is available on GitHub (https://github.com/thomasball42/pvm_curve_modelling) and in a permanent repository (DOI: 10.5281/zenodo.14419062).

**Acknowledgements**

The authors thank Ian Bateman for his comments on this manuscript, and Michael Dales and Anil Madhavapeddy for their support of our use of the Cambridge University Department of Computer Science and Technology EEG group high performance computing facilities. TSB was funded by UK Research and Innovation's BBSRC through the Mandala Consortium and (grant number BB/V004832/1) and UK Research and Innovation's cross research council responsive mode (CRCRM) pilot scheme (grant number MR/Z505456/1). BB gratefully acknowledges the support of the Dragon Capital Chair on Biodiversity Economics funding his previous postdoc position. AB is grateful for grants from the Tezos Foundation and Tarides to the Cambridge Centre for Carbon Credits (grant code NRAG/719).

# Supplementary Materials

## A. Structure and assumptions of the population simulation models

### Overview of methods

We devised a series of population simulations with which to explore the relationship between population carrying capacity and probability of extinction. Each of our models is based upon a logistic growth model, in which the growth rate at a given time is based upon the number of current individuals in the population $N$ over the carrying capacity $K$, and the maximum growth rate $r_{\max}$, such that realised growth rate is maximised when $N = 1$ and is zero when $N = K$. For a mathematical formalisation of this and the following models, see the latter half of this section. We set the time step for our simulations to be one year, since empirical species data, for example survival rates and breeding rates, are typically expressed on annual timescales.

In Model A, the growth rates of the two sexes are calculated independently of each other. To represent demographic stochasticity, at each time step the new number of individuals of each sex was drawn from a Poisson distribution. Environmental stochasticity was implemented as a random variable normally distributed about zero, with standard deviation $\sigma$. We reasoned that environmental factors affecting one sex would probably affect the other equally, so we applied the same environmental stochasticity to both sexes at each time step. Model B is similar to Model A but introduces demographic dependence such that the growth rate of each sex is dependent on the number of individuals in both sexes combined: if there is a reduced number of males at a given timestep, the growth rate of females would also be affected.

Model C builds on Model B but introduces more complex features of demographic stochasticity: biparental care, and a representation of an age at first breeding (a time to maturity). The growth rate of each sex is now governed by the number of adult males and females in the population at some time before the current time step, determined by the age at first breeding. This necessitated a reformulation of the model, and the inclusion of additional parameters $S_a$, the annual survival rate of adults in the population, and $B$, the age at first breeding. Combinations of $S_a$ and $r_{\max}$ give plausible values of $B$, as described in Supplementary Appendix B.

Model D is an extension of Model C incorporating a representation of temporally-autocorrelated environmental stochasticity. We replaced the normally distributed random variable from the previous models with a random walk with a central tendency, such that the realisation of environmental noise at a given time was more likely to be closer to that of the previous time step. This necessitated the inclusion of a final model parameter $Z$, which is a 'reversion' factor: the tendency of the noise to 'walk' back to zero. For $Z = 0$, the environmental noise is effectively a random walk with no tendency to revert to zero, and for $Z = 1$, the environmental noise is effectively drawn from a normal distribution about zero at each timestep (as is the case in Models A, B, and C).

For each model, we ran simulations across a range of parameters, varying $r_{\max}$ between 0.055 and 0.774, $\sigma$ between 0.05 and 0.55, $S_a$ between 0.35 and 0.95, and $Z$ between 0.01 and 1.0. For the rationale behind the selection of these parameter spaces see Supplementary Appendix B. For each set of inputs, we ran simulations for a range of $K$ values, which were geometrically distributed between 1 and 3 million. For each individual value of $K$ we ran the model for 10,000 repeats, recording the probability of extinction $P_E$ as the proportion of those repeats that reached extinction at or before 100 years. The modified Gompertz model was fitted to the results for each set of input parameters following a process described in Supplementary Appendix C.

In the basic runs for Models A through D, the simulation was initialised with the number of individuals equal to the carrying capacity $K$. To explore the effect of beginning with populations not at the carrying capacity, we initialised Model A with $r_{max}$=0.56 and $\sigma$=0.15, varying the starting population $N_0$ in two ways. First, such that $N_0$ was proportional to $K$ in 8 increments between 0.125 and 1.0, second such that $N_0$ was a constant, regardless of the value of $K$, in incremental values between 1 and 1 million in factors of 10. As before, the number of individuals in each sex was set at half the total carrying capacity. We then varied $K$, running 10,000 repeats for each value as before, and examined the relationship between $P_E$ and $K$.

## Formulation and simulation

We wished to examine the relationship between the probability of extinction of a population within a defined time period and its carrying capacity, and to assess how this shape is affected by life-history characteristics of the species and the level of environmental stochasticity it is subject to. To do this, we constructed four population viability models of increasing complexity and ran simulations for a range of their input parameters. In each model, the number of individuals in the population following time step $dt$ is governed by the number of individuals at the previous time step $t$, and the population growth rate during that time step by the following relationship:

$$N_{t+dt} = N_t + r_t * N_t \tag{A1}$$

where $N_t$ is the number of individuals at $t$ and $r_t$ is the growth rate at $t$. The models are characterized by different forms of the growth rate $r$ and their realisation of environmental stochasticity. Importantly, in all models discussed here, the realised value of $N_{t+dt}$ is an integer from a Poisson distribution with an expectation value of $N_{t+dt}$. This discretization represents a source of demographic stochasticity in all of our models.

**Model A**

The simplest realisation of the growth rate $r$ in our models is a logistic density dependent growth rate, with some environmental stochasticity, the equation for which is as follows:

$$r_t = r_{\max} * \left(1 - \frac{N_t}{K}\right) dt + Q_t, \tag{A2}$$

where $r_{\max}$ is the maximum growth rate of the population, $K$ is the carrying capacity of the population, and $Q_t = Q\sqrt{dt}$, $Q \sim N(0, \sigma)$ is the environmental stochasticity term, which is drawn from a normal distribution with mean zero and with standard deviation $\sigma\sqrt{dt}$. This formulation includes, but does not separately identify, some of the mechanisms for density dependence affecting demographic rates, such as influences on survival and age-specific fecundity caused by competition for food, predation patterns, and disease. We adopted annual timescales because many sources of environmental stochasticity, such as the effects of weather, vary substantially from year to year. Moreover, for most species breeding cycles occur annually, and hence data on growth rates is typically expressed as annual. Thus, following the application of a timestep of one year (i.e. $dt = 1$), the annual growth rate of the population becomes:

$$r_t = r_{\max} * \left(1 - \frac{N_t}{K}\right) + Q_t, \tag{A3}$$

for which $r_\text{max}$ and $Q_t$ are defined as the maximum annual growth rate, and annual stochastic variation respectively.

Model A as described so far considers only one sex, so we adapt it into a two-sex model of a sexually-reproducing species by considering the total population as two distinct populations of males and females with separate carrying capacities $K_m$ and $K_f$, each equal to half total $K$. In Model A, male and female growth rates are independent of each other, but at each time interval share the same realised environmental stochasticity as captured by $Q_t$ (i.e. both sexes are subject to the same environmental pressures). Equations (A1) and (A3) together generate expected real numbers of males and females at each time step, which are then converted to integer values via the Poisson process. This is done separately for males and females and hence the number of individuals of each sex are able to deviate from each other over time.

## Simulation setup for Model A

To model a population with given values of $r_\text{max}$ and $\sigma$, we initialized the population at its carrying capacity with an equal number of males and females such that at $t = 0$:

$$N_{t=0,\text{males}} + N_{t=0,\text{females}} = K_\text{males} + K_\text{females} = K \tag{A4}$$

We then ran the simulation for 100 time intervals (100 years). If at any point the population of either males or females (or both) reached zero, the population was considered to be extinct. We repeated this process for 10000 separate populations and then calculated the probability of extinction $P(E_{t \leq 100})$ as the proportion of those 10000 runs that went extinct within 100 years. We did this for various combinations of $r_\text{max}$ and $\sigma$. The choice of values for those parameters is discussed below in the choice of input parameters section (Appendix C). We carried out this process for integer $K$ values geometrically distributed between 1 and 3 million, since we are interested in the shape of the relationship between $P_E$ and $K$.

## Model B

Next, we adapted Model A such that the growth rates for males and females were no longer independent of each other, instead being dependent on a density-dependent relationship of *r* to the total number of males and females combined; we call this Model B. The growth rate in Model B is described as follows:

$$r_{t,\text{sex}} = r_\text{max} * \left(1 - \frac{N_{t,\text{males}} + N_{t,\text{females}}}{K_\text{males} + K_\text{females}}\right) + Q_t, \tag{A5}$$

where $r_{t,\text{sex}}$ is the growth rate at $t$ for the given sex. As before, $Q_t$ is normally distributed with mean zero and standard deviation $\sigma$, and applied to both sexes in the same way at each time step. Note that this means that at a given time step the growth rate for both sexes is the same. We ran Model B simulations across the same parameter space as Model A.

## Model C

For the next model, we began to introduce more complex population processes. Unlike models A and B, in which demographic stochasticity is governed only by differences in the outcomes of the Poisson processes in generating integer numbers of individuals from real numbers, Model C includes an additional source of demographic stochasticity arising from monogamy and biparental care. In this model, the fecundity of the population is reduced in the presence of a disparity between the populations of each sex arising from the stochastic processes.

Conceptually, this model aims to represent species in which individuals require long-term access to at least one member of the opposite sex during the breeding season to reproduce successfully.

To implement this, firstly we defined a modified growth rate $r'_t$ in the following way:

$$r'_{t,B} + 1 = S_a + V_{t,B} \tag{A6}$$

here, each side of the equation is effectively the population multiplier between time $t$ and $t + dt$; in models A and B, this was simply $r_t + 1$: the existing population plus a per-adult growth rate. We then reformulated this multiplier by the introduction of $S_a$: the annual probability of survival for an adult in the population, and $V_{t,B}$: the number of 'recruits' to the adult population *per adult* at time $t$, where $B$ is the age at recruitment to the adult population - i.e. a time to maturity, or age at first breeding. To account for the potential mismatch between the number or males and females at time $t - B$ and consequent impact on fecundity, we defined $V_{t,B}$ in the following way:

$$V_{t,B} = \frac{N_{t-B}}{N_t} * (r_{t-B} + 1 - S_a) * \frac{\min(N_{m,t-B}, N_{f,t-B})}{\max(N_{m,t-B}, N_{f,t-B})} \tag{A7}$$

where $N_{m,t-B}$ and $N_{f,t-B}$, and $r_{t-B}$ are the number of males and females, and growth rates as calculated by either model A or B, at time $t - B$ respectively. Putting equations A6 and A7 together, and adding annual environmental stochasticity $Q_t$ as before, we get the modified growth rate:

$$r'_{t,B} = S_a - 1 + \frac{N_{t-B}}{N_t} * (r_{t-B} + 1 - S_a) * \frac{\min(N_{m,t-B}, N_{f,t-B})}{\max(N_{m,t-B}, N_{f,t-B})} + Q_t. \tag{A8}$$

For the sake of demonstration, here we combine equations (B1) and (B8):

$$N_{t+dt} = S_a N_t + N_{t-B} * (r_{t-B} + 1 - S_a) * \frac{\min(N_{m,t-B}, N_{f,t-B})}{\max(N_{m,t-B}, N_{f,t-B})} + N_t Q_t. \tag{A9}$$

Broken down by term, the number of individuals at $t + dt$ is the number of existing adults expected to survive, plus the number of adolescent individuals reaching adulthood (modified for any sex disparity at time $t - B$), plus the environmental stochasticity term affecting any noise in the number of adults: excess deaths or excess arrivals.

The values of $S_a$, $B$, and $r_{\max}$ are related to each other in the following way, as can be obtained by rearrangement of Equation 17 in Niel and Lebreton (2005):

$$B = \frac{1}{r_{\max}} - \frac{S_a}{e^{r_{\max}} - S_a} \tag{A10}$$

to allow the calculation of $B$ values from combinations of $S_a$ and $r_{\max}$. Using this approach, we then ran simulations for Model C across the same parameter space as used previously, but now introducing the new parameter $S_a$. When running simulations of the pre-existing parameter space for different values of $S_a$ the total parameter space can be become unmanageably large, so we used values for $S_a$ between 0.35 and 0.95 in increments of 0.15. This range is based upon a compilation of estimated annual survival rates of birds provided by Bird et al (2020).

## Model D

In Models A, B, and C, $Q$ was drawn each time step from a normal distribution centred at zero (standard deviation of $\sigma$) independent from previous values of $Q$. In the final model we introduce temporal autocorrelation to the environmental stochasticity term. This is motivated by the real-world observation that periods of adverse environmental conditions tend to be clustered rather than fluctuate independently from year to year, as we assumed in Models A-C (Green et al. 2020). To do this, we took the simple approach of having the environmental noise 'random walk' at each time step, with a tendency to return to a central value of zero. The implementation of this process is as follows:

$$Q_{t+1} = (1 - Z) * Q_t + \varepsilon_t \qquad (A11)$$

where $Q_t$ is the realised value of the environmental stochastic noise at time $t$, $Z$ is a reversion factor: the tendency of the process to move toward zero. $Q_t$ is the realised value of stochastic noise at the previous timestep (set at zero for the first timestep), and $\varepsilon_t \sim N(0, \sigma)$ is a white noise process with standard deviation $\sigma$, as before. Choosing Z=1 yields the normally distributed stochastic increment used in Models A, B, and C.

We combined the growth rate model from Model C with this new realisation of environmental stochasticity to form Model D. We conducted simulations of Model D across the previous parameter space, but now introduced the new parameter $Z$, which we ran for values of 0.01, 0.25, 0.5, and 0.75.

## B. Choice of input parameters

To generate plausible values for $r_{\max}$ we applied the demographic invariants method (DIM), which is based upon a study by Charnov (1993). Equation 17 in Niel and Lebreton (2005) describes the DIM relationship as follows:

$$e^{r_{\max}} = \exp\left[\left(B + \frac{S_a}{e^{r_{\max}} - S_a}\right)^{-1}\right] \qquad (B1)$$

where $B$ is the age at first breeding, and $S_a$ is the annual probability of adult survival. Using data on the values for $B$ and $S_a$ for all ~10,000 of the world's bird species listed in the supplementary materials of Bird et al. (2020) we recursively solved equation B1 for each species to generate a range of plausible values for $r_{\max}$. We took the 10th and 90th percentiles of the calculated values to bound our parameter space and used 15 values of $r_{\max}$, linearly spaced between 0.055 and 0.774. For the standard deviation of the environmental stochasticity term $\sigma$, we chose to use $\sigma$ values between 0.05 and 0.55 in intervals of 0.03. The upper limit of $\sigma = 0.55$ was sufficient to capture all valid runs, for reasons described in the analysis section.

Overall, our choices of $\sigma$ and $r_{\max}$ yield 255 parameter combinations, meaning that we ran a total of just over 450 million individual simulations for model A. All models were written in Python (the underlying code for which is available at: https://github.com/thomasball42/pvm_curve_modelling).

Some of these combinations of $\sigma$ and $r_{\max}$ we considered to be invalid within our modelling framework. If $\sigma$ is too large compared to $r_{\max}$, population growth is never able overcome environmental stochasticity, leading to an unstable population that is always at risk of extinction regardless of the carrying capacity $K$. We are able to derive this relationship analytically in the case of single-sex, parthenogenic populations (see appendix E), but the inclusion of an additional sex and Poisson process complicates the derivation, thus we consider this empirically.

We regarded combinations of input parameters that lead to this effect to be unrealistic within our framework; our models aimed to explore the relationship between carrying capacity and extinction risk over medium timescales for stable populations. Given that currently extant species typically originated millions of years ago and usually at least hundreds of thousands of years ago (Barnosky et al. 2011), a population for which $P_E > 0$ within 100 years at large K would be likely to have gone extinct by the present day. We found that this phenomenon occurred by or before $\sigma = 0.55$ for all models and values of $r_{\max}$, hence our choice of this upper limit for this input parameter.

To remove these runs from our results we simply excluded runs with any parameter combinations for which there were no data points at which $P_E = 0$ for any value of K up to 3 million. With this approach, it is possible that we exclude some simulation sets that might reach $P_E = 0$ given an even higher carrying capacity than we took to be our upper limit. However, by inspection of our result-space we posit this this may only have occurred for some Model D runs with $Z = 0.01$. For practical purposes, the majority of simulations express clear asymptotic behaviour with an asymptote at $P_E = 0$ well before a carrying capacity of 3 million is reached, so we think that these exclusion criteria are appropriate.

## C. Fitting functions describing the shape of the relationship of $P$ to K

To fit functions relating $P_E$ to $K$ for our selected combinations of input parameters we used the optimize.curve_fit function from the Scipy Python library (Virtanen et al. 2020), which attempts to find the optimal parameters to fit a chosen function to the provided data. However, we found that the fitting process for the modified Gompertz model (main text Equation 1) was very sensitive to initial values of $\gamma$, which we describe as the 'shape parameter' of the curve. It was thus necessary to provide approximate starting values to get the best fit. We achieved this by transforming the function as follows:

$$\ln[-\ln(1 - P_E)] = a + b * K^\gamma \tag{C1}$$

where symbols have their previous meanings. For a range of $\gamma$ values between 0 and 5 in increments of 0.05 we tested for linear correlation between $\ln[-\ln(1 - P_E)]$ and $K^\gamma$; the initial $\gamma$ value that achieved the best coefficient of determination was then used to seed the fitting optimisation function along with the corresponding values for $a$ and $b$. As is described in the main text this allowed to us to achieve fits with $r^2$>0.999 for a large portion of the curves.

## D. Implementation of Allee effects

**Description of Allee effect formulation and modelling**

To test whether the inclusion of an Allee effect in our simulation changed the generality of our result, we modified the formulation of the growth rate in Model A to include an Allee effect term. Populations in a mean-reverting growth system with an Allee effect have two equilibria: a stable equilibrium at carrying capacity K (as is the case in the basic logistic growth model) and an additional, unstable equilibrium at population $\theta$, where $\theta$ < K. To do this, we used the formulation in Dennis (2002), modifying the growth rate equation (A2), adding an additional term as follows:

$$r_t = r_{\max} * \left(1 - \frac{N_t}{K}\right) - \frac{\upsilon\theta}{\theta + N_t} + Q_t, \tag{D1}$$

where $\upsilon$ is the strength of the Allee effect, and other symbols have their previous meanings. The strength parameter $\upsilon$ can take any positive value and is somewhat analogous to the growth rate

$r_{max}$ – where $r_{max}$ determines the strength of the attractive force of the stable equilibrium at $K$, $v$ governs the strength of the repelling force of the unstable equilibrium at $\theta$.

To test the effect of Allee effects in our simulations, we varied both parameters in turn, repeating each combination of $\theta$, $v$, and $K$ 10,000 times as in our other simulations, and calculating the probability of extinction as the portion of the 10,000 population that went extinct within 100 time steps. The results of these simulations are shown in Figures D1 and D2 below.

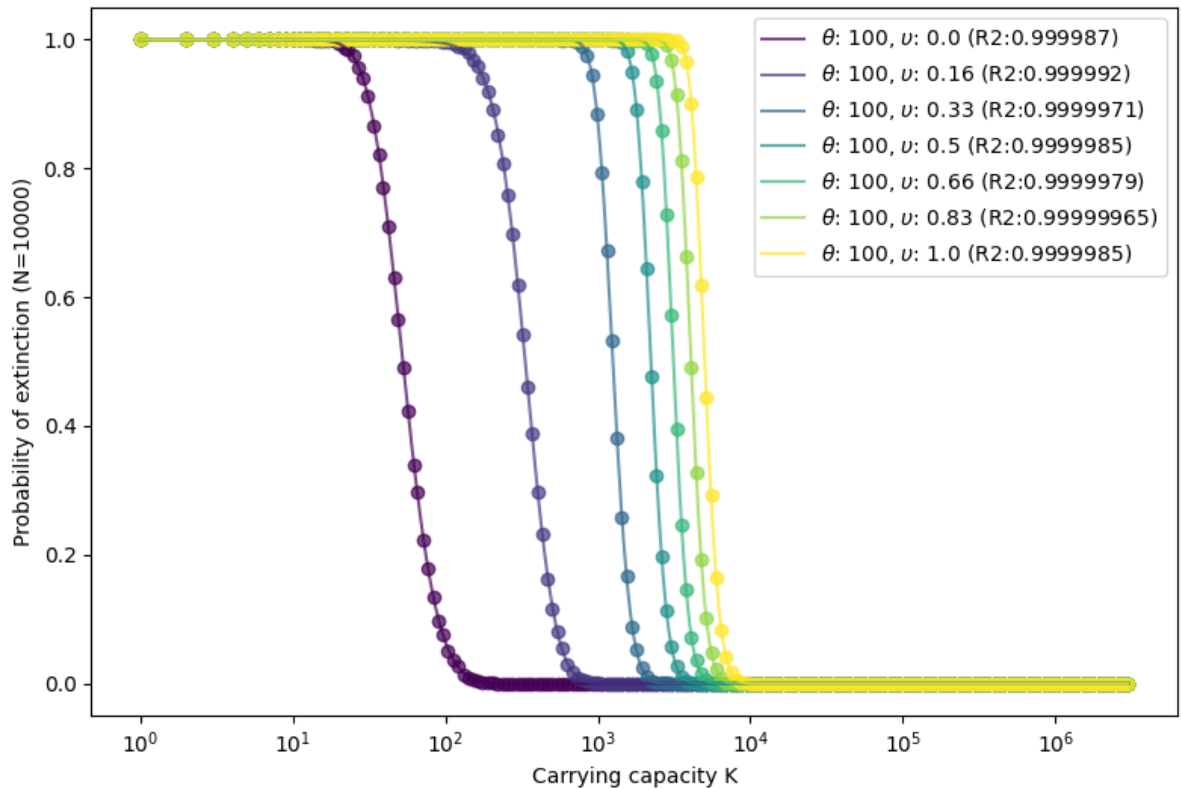

Figure D1. Probability of extinction within 100 time steps $P_E$ versus carrying capacity $K$ across 10,000 identical populations in the presence of an Allee effect when varying the Allee effect strength parameter $v$ between 0 and 1. Simulated $P_E$ values are represented by circles, best-fit modified Gompertz curves are shown as lines, with their corresponding $v$ and $r^2$ values shown in the legend. The Allee effect critical population threshold $\theta$ was set at 100 for all simulations.

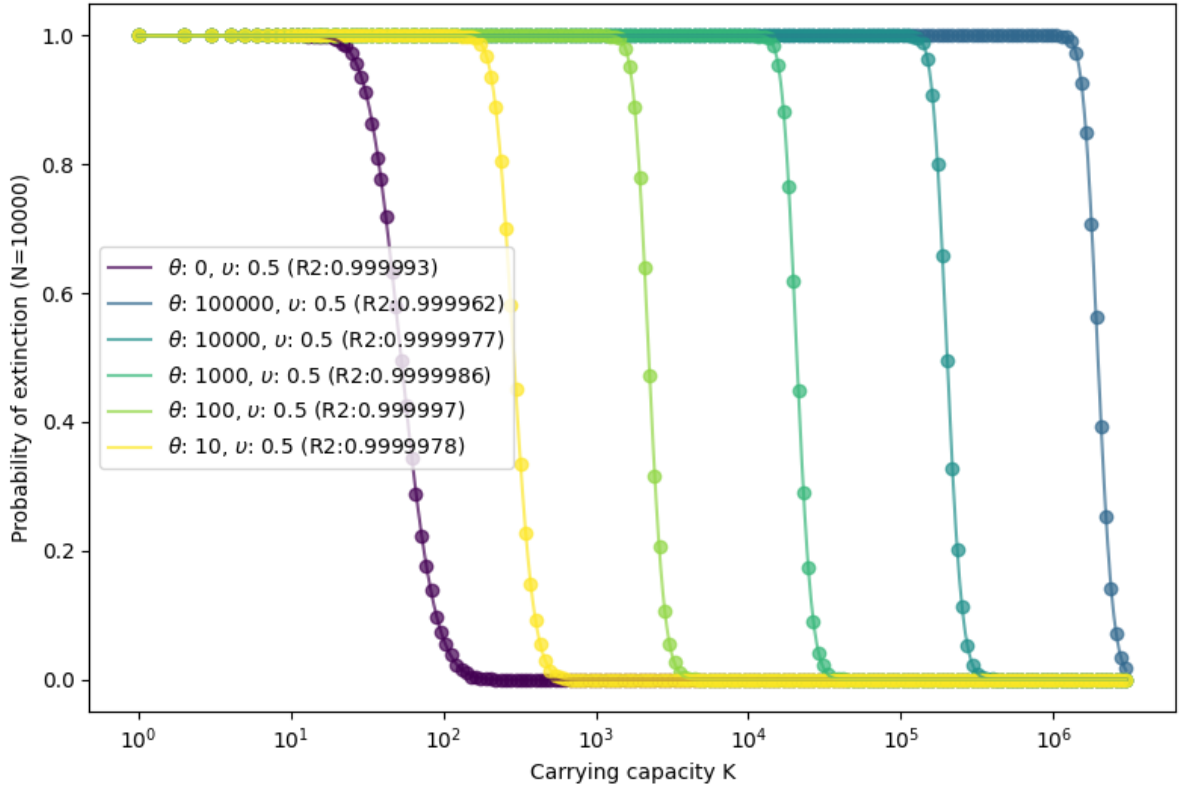

Figure D2. Probability of extinction within 100 time steps $P_E$ versus carrying capacity $K$ across 10,000 identical populations in the presence of an Allee effect when varying the Allee effect critical population threshold $\theta$ between 0 and 100000. Simulated $P_E$ values are represented by circles, best-fit modified Gompertz curves are shown as lines, with their corresponding $\theta$ and $r^2$ values shown in the legend. The Allee effect strength parameter $\upsilon$ was set at 0.5 for all simulations.

## E. Fitting the modified Gompertz to empirical examples

**1. Empirical study of extinction rates observed in isolated populations of bighorn sheep**

A study reported local extinctions observed in isolated populations of bighorn sheep (*Ovis canadensis*) in the western USA. We obtained data from Figure 1 of Berger (1990) and its legend on the numbers of isolated populations that had persisted or gone extinct during periods of 10, 20, 30 and 40 years after the beginning of the observation period. Results were also reported for populations after 50, 60 and 70 years, but monitored populations for those intervals were too few and with too restricted variation in extinction rates to permit reliable fitting of the extinction functions. The initial size of each population was categorised as 1-15, 16-30, 31-50, 51-100 and >100. We took the means of the bounds of these categories to be equivalent to carrying capacity K for each group of populations and assigned the >100 category a K value of 101. These initial population sizes may not correspond to K precisely, but our conclusions about the form of the relationship of probability of extinction to K would not be affected if K was a fixed multiple of initial population size. We obtained the number of populations that had gone extinct *n* and the number that persisted *m* for each time interval and each initial population size class. For each time interval, we initially modelled the probability of a population having gone extinct $P_E$ in relation to K as the modified Gompertz function:

$$P_E(K) = 1 - \exp(-\exp(a + b\,K^\gamma)). \tag{E1}$$

We used an iterative maximum-likelihood method to estimate parameters a, b and γ. We calculated, for a given monitoring time period and for each initial population size class, the quantity

$$n * \ln(P_E{'}) + m * \ln(1 - P_E{'}), \tag{E2}$$

where $P_E{'}$ is the value of $P_E$ calculated assuming preliminary values of a, b and γ. We summed the calculated quantity across the initial population size classes for a given time interval to give the log-likelihood of the observed data for that time interval, given the provisional parameter estimates. At each subsequent iteration, we used a quasi-Newton algorithm to increase the log-likelihood of the data, given the values of a, b and γ. The procedure continued until the log-likelihood stabilised at a maximum value. A separate set of model parameter values was estimated for each time interval.

This procedure initially gave unrealistic results because some analyses gave values of $P_E{'}$ that were less than 1 when K was 1. A population with a carrying capacity of 1 is almost certainly doomed to extinction within even the shortest time period for which the sheep populations were monitored (10 years). We therefore modified Equation E1 to

$$P_E = 1 - \exp\left(-\exp\left(1.52718 + b * (K^\gamma - 1)\right)\right) \tag{E3}$$

which constrains $P_E{'}$ to be 0.99 when $K = 1$. We considered this approximation sufficient to make our fitted models realistic. We used Equation E3 to fit separate modified logistic functions for each of the four periods over which extinctions were monitored (10, 20, 30 and 40 years). It was necessary to estimate two parameters (b and γ) for each of the four time periods separately, so a total of eight estimates was required. We also fitted a simplified version of this model in which the parameter b was treated as a linear function of the time $t$ over which extinctions were monitored. This gave:

$$P_E = 1 - \exp\left(-\exp\left(1.52718 + (f + g * t) * (K^\gamma - 1)\right)\right) \tag{E4}$$

This formulation assumes that b is a linear function of t with intercept f and slope g and that the shape parameter γ is the same for all four time periods. Hence, the model of data from all four time periods required the estimation of only three parameters (f, g and γ). We used the maximum-likelihood procedure described above to fit all of the models. We refer to this formulation as the 'all-periods' version of the model.

Observed and modelled values of $P_E$ in relation to $K$ are shown in Figure E1a for each of the time intervals 10, 20, 30 and 40 years for the version of the model parameters specific to each time period (Equation 3). For this model with time-period-specific parameters, the square of the Pearson correlation coefficient ($r^2$) between the 20 observed and expected pairs of $P_E$ values for all four time periods was high and the root-mean-square of differences (RMSD) between observed and modelled $P_E$ values was quite small (Table E1). In addition, all of the modelled values of $P_E$ for a given $K$ lay within the Clopper-Pearson 95% confidence intervals (Clopper & Pearson 1934) for the individual values of $P_E$ obtained from the data (Figure E1a; Table E1). Our modification to the Gompertz function required the inclusion of the shape parameter γ, where γ <> 1. We wished to assess the effect of this modification on the fit of the model, so we fitted a version of the function shown in Equation E3 with γ set to 1, which is equivalent to not having a shape parameter. We refer to this as the unmodified Gompertz function. The $r^2$ value for the unmodified Gompertz model with time-period-specific parameter estimates of b was lower and the RMSD was much larger. A smaller proportion (11 (55%) of the 20 of the modelled values for a given K, cf. 100%) lay within the Clopper-Pearson 95% confidence interval for the value of $P_E$ obtained from the data (Table E1).

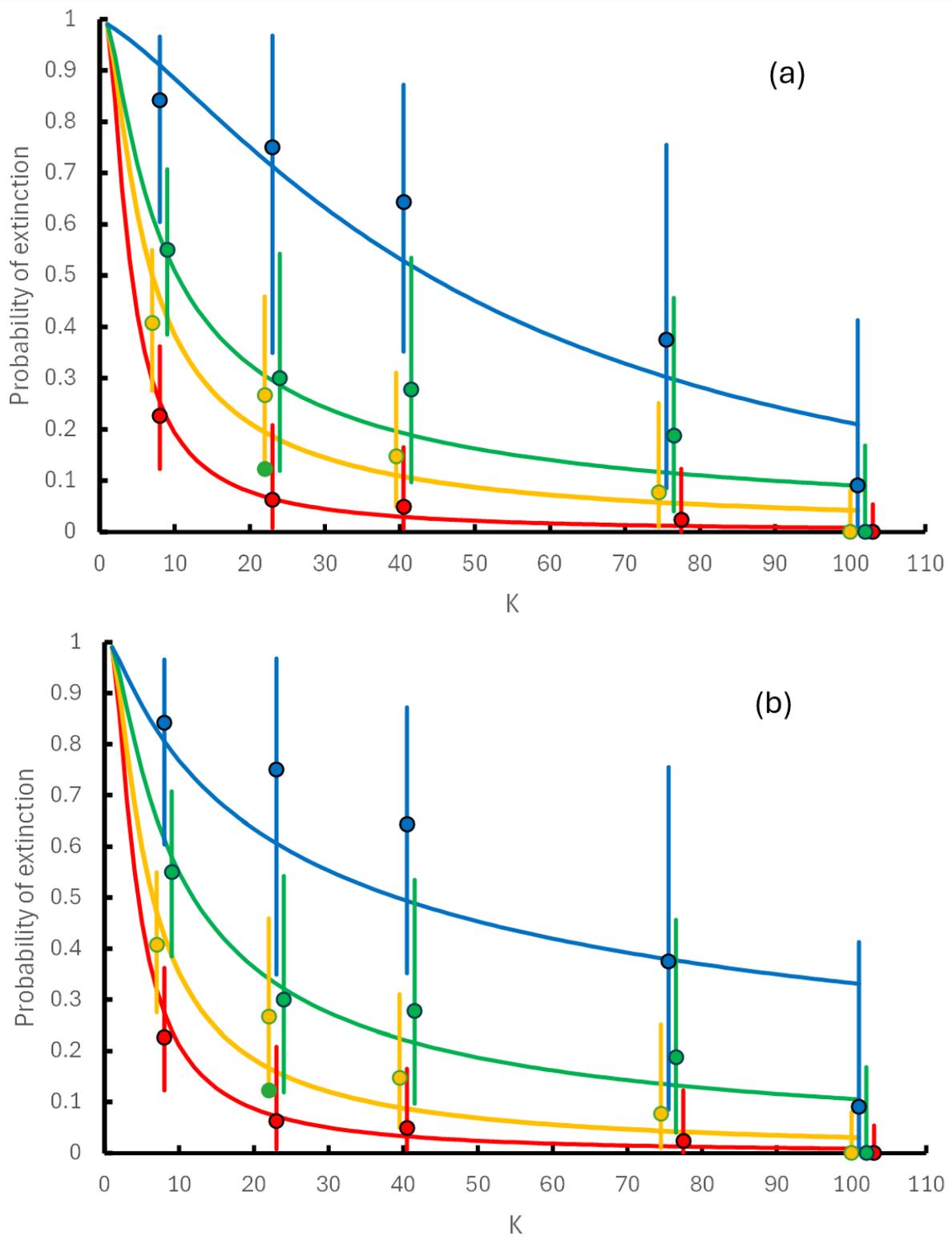

Figure E1. Probability of extinction of groups of populations of bighorn sheep within different specified time periods, in relation to their initial population size K, assumed to be equivalent to carrying capacity. Observed proportions (circles) are from Figure 1 of Berger (1990) with vertical lines showing their 95% binomial confidence intervals (Clopper & Pearson 1934). Colours denote the four time intervals over which extinction rate was assessed: red = 10 years; gold = 20 years; green = 30 years; blue = 40 years. Fitted modified Gompertz functions are shown for each time interval, with the same colour coding, for (a) the model with time-period-specific parameter estimates and (b) the all-periods model with parameter b assumed to be a linear function of time period duration t. In this model the shape parameter *Y* is assumed to be the same in all four time periods.

Table 1. Performance of modified and unmodified Gompertz models in fitting data from three studies of extinction probability $P_{ext}$ in relation to carrying capacity. The square of the Pearson correlation between observed and modelled probability $P_{ext}$ values ($R^2$), the root-mean-square-difference (RMSD) between these values and the percentages of 20 modelled $P_{ext}$ values that lay within the 95% Clopper-Pearson confidence interval (CI) of the observed value of $P_{ext}$ are shown. CI of individual observed values could only be calculated for the bighorn sheep study, so NA is given in these columns for the other two studies.

| Species | $R^2$ | | RMSD | | Percentage of expected within CI of observed $P_{ext}$ | |
| --- | --- | --- | --- | --- | --- | --- |
| | Modified | Unmodified | Modified | Unmodified | Modified | Unmodified |
| Bighorn time-period-specific | 0.9429 | 0.8069 | 0.0607 | 0.1713 | 100 | 55 |
| Bighorn all-periods | 0.8879 | 0.7941 | 0.0850 | 0.1771 | 100 | 55 |
| NZ brown mudfish (low) | 0.9948 | 0.9948 | 0.0294 | 0.0294 | NA | NA |
| NZ brown mudfish (high) | 0.9815 | 0.9813 | 0.0558 | 0.0561 | NA | NA |
| Grizzly bear | 0.9993 | 0.9992 | 0.0115 | 0.0121 | NA | NA |

Observed and modelled values of $P_E$ in relation to K are shown in Figure 1(b) for the all-periods version of the model, using the modified Gompertz function as described by eq.(4). The $r^2$ for this model was high and the RMSD was low, but these measures of fit were less good than for the time-period-specific model, as would be expected for a model with fewer fitted parameters. (Table 1). However, all of the modelled values for a given K again lay within the Clopper-Pearson 95% confidence intervals for the value of $P_E$ obtained from the data (Figure 1(b); Table 1). The $R^2$ value for the unmodified Gompertz all-periods model was considerably lower and the RMSD was much larger than for the modified version. Eleven (55%) of the 20 of the modelled values for a given *K* lay within the Clopper-Pearson 95% confidence interval for the value of $P_E$ obtained from the data (Table E1). We conclude that the modification of the Gompertz function to include the shape parameter *Y* resulted in a substantial improvement of the fit of the model to the bighorn sheep data, compared with the unmodified Gompertz and that the modified Gompertz model fits the observations of this empirical study of observed extinction rates of bighorn sheep reasonably well.

## 2. Modelling study of the probability of extinction of populations of New Zealand brown mudfish

White et al. (2017) used the population viability modelling software package RAMAS METAPOP to model the probability of extinction within 500 time steps of simulated populations of a New Zealand fish (*Neochanna apoda*) in relation to carrying capacity *K*. We took results from the points plotted on their Figure 4 and fitted the modified Gompertz model to them using nonlinear least squares. We did this for their two sets of simulations that assumed low or high environmental stochasticity caused by droughts. Our model fitted results from both sets of simulations well (Table E1). There were only slight differences the fit of the modified and unmodified versions of the Gompertz function (Table E1). We conclude that both the modified and unmodified Gompertz models fit these simulated extinction rate results obtained from a frequently-used population viability approach reasonably well.

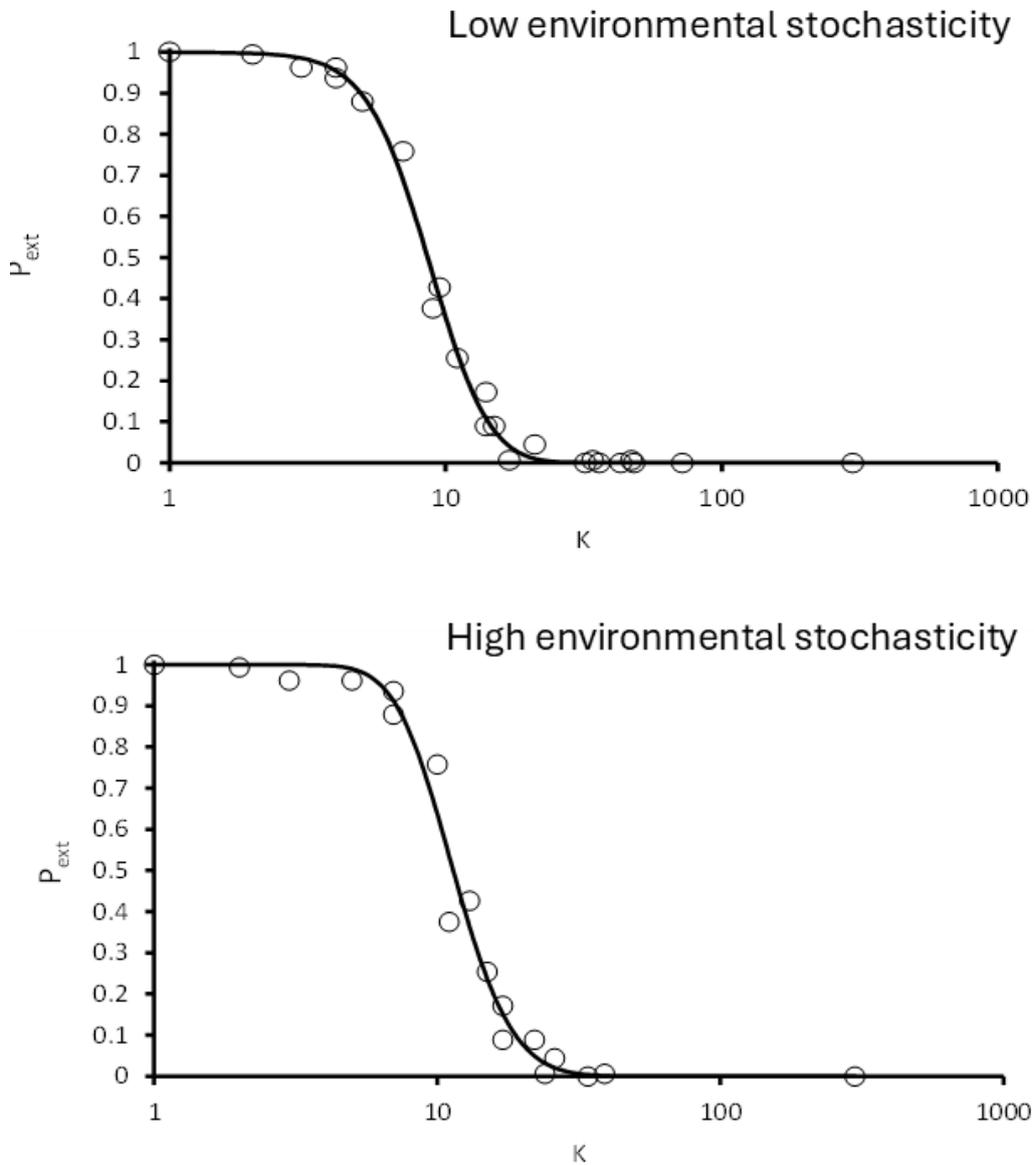

Figure E2. Probability of extinction within 500 time steps of simulated populations of the New Zealand brown mudfish *Neochanna apoda* in relation to carrying capacity K. Plotted points are results from simulations and curves are fitted modified Gompertz models. The upper panel is for low environmental stochasticity caused by droughts and lower panel is for high stochasticity.

**3. Study of the probability of extinction of grizzly bears in Yellowstone Park simulated using a demographic model with parameters based upon field studies**

Shaffer & Samson (1985) estimated the probability of extinction of simulated populations of grizzly bears *Ursus arctos* in Yellowstone National Park at various different assumed carrying capacities. The simulation model they used was a discrete time, discrete number formulation employing sex and age structure, mortality and reproductive rates, and density-dependent relationships revealed by an independent analysis of Craighead et al's (1974) 12 years of data on the Yellowstone grizzly bear population (Shaffer 1978). Demographic stochasticity was introduced via the discrete number (integer) formulation of the model. Integer individuals are created and decisions about their survival and reproduction are made through use of pseudorandom number generators. Environmental stochasticity was introduced via the use of the variance in the empirical estimates of mortality and reproductive rates. Extinction was defined as the death of the last adult. The key results, from our point of view, of Shaffer & Samson's study are presented in their Table E1. We fitted the modified Gompertz model to those results using nonlinear least squares. Figure E3 shows the values from Shaffer & Samson and the fitted model. Both the modified and unmodified Gompertz functions fitted the results well (Table E1). We conclude that both of these Gompertz models fit the simulated extinction rate results obtained from this population viability approach reasonably well.

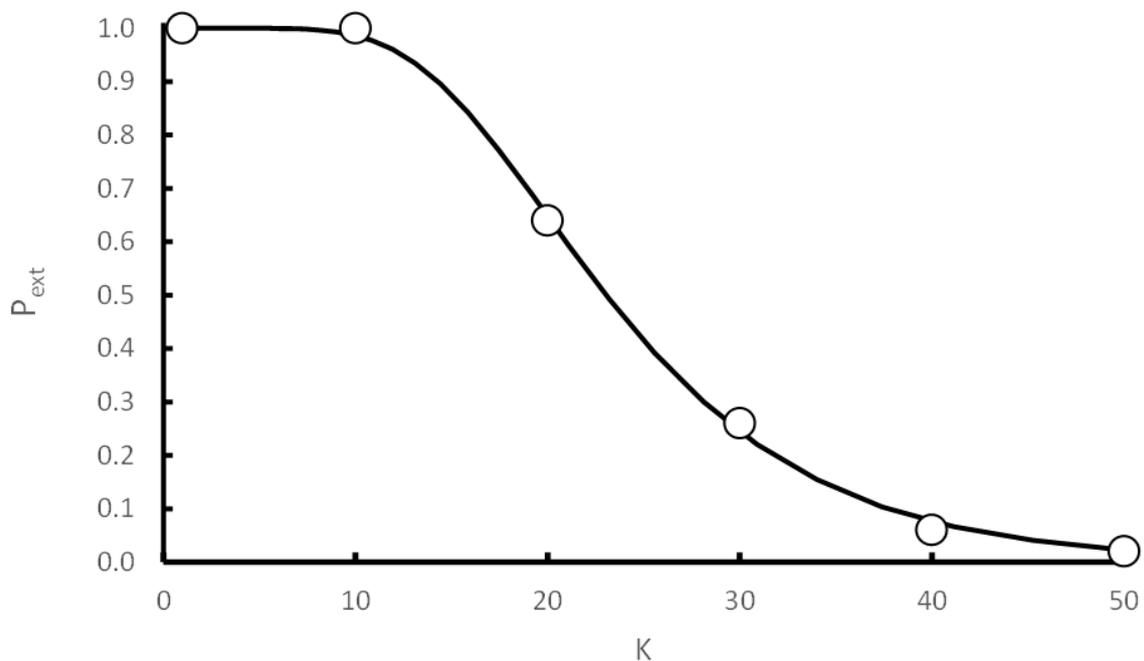

Figure 3. Probability of extinction in a 100-year period $P_{ext}$ of grizzly bears *Ursus arctos* in relation to carrying capacity K. Results (circles) are from a simulation model in which demographic rates of bears were assumed to vary among years according to upon empirical observations of the variance of environmental stochasticity in demographic rates and assumptions about demographic stochasticity. The curve shows the modified Gompertz model fitted using nonlinear least squares.

**Comparison of the performance of the modified and unmodified Gompertz models**

Our analyses of these three examples indicate that the modified Gompertz function fits the data from all three studies well. The modified Gompertz performed substantially better than the unmodified version for one of the species, the bighorn sheep. Figure 4, provides insights into this difference among the studies by plotting the estimated values of the shape parameter $Y$ from all of the modified Gompertz analyses. It can be seen that $Y$ is less than 1 for seven of the eight estimates. All of the $Y$ estimates from the bighorn sheep study are substantially less than one, as expected from the less good fit of the unmodified compared with the modified model for this species. For the brown mudfish and grizzly bear studies, the estimates of $Y$ were much closer to 1 and the fits of the modified and unmodified models were therefore similar for these two species. Although no firm conclusions can be drawn from this small sample of studies, the tendency for $Y$ to be less than 1 is strikingly similar to the same tendency we observed for fitted $Y$ values for Models A – D from our simulation results (see Section G. and Figure G6).

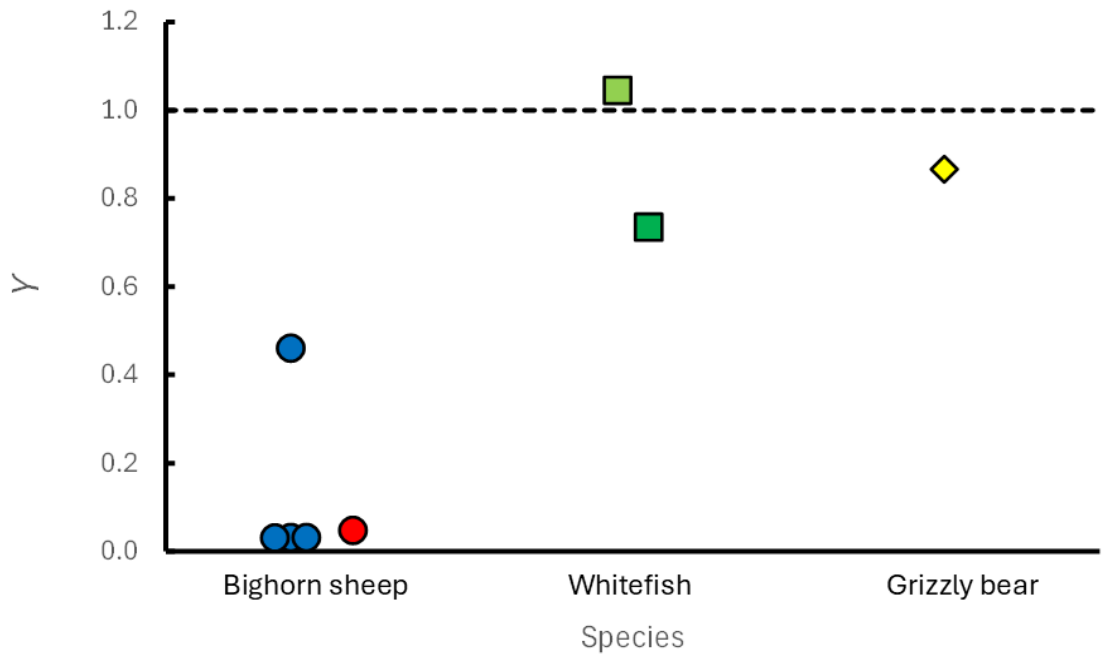

Figure E4. Estimated values of the shape parameter $\gamma$ estimated by fitting the modified Gompertz to data from studies of bighorn sheep, New Zealand brown mudfish and grizzly bears. For bighorn sheep, the blue circles show four values from the time-period-specific model and the red circles shows the value from the all-periods model, For New Zealand brown mudfish the light green square shows the value for the low stochasticity model and the dark green square shows the value for the high stochasticity model. The dashed horizontal line represents a $\gamma$ value of 1, which is the value of the shape parameter for the unmodified Gompertz model.

## F. Theoretical basis for the modified Gompertz

The continuous time counterpart of a population process $X_t$ can be expressed as the logistic stochastic differential equation given by

$$dX_t = rX_t\left(1 - K^{-1}X_t\right)dt + \sigma X_t\, dW_t, \qquad X_0 = x_0, \tag{F1}$$

for all times $t \geq 0$, where $r$ is the growth rate, $K$ is the carrying capacity and $W_t$ is the standard Brownian motion, here calibrated with a geometric diffusion coefficient $\sigma X_t$. Note that (F1) can be interpreted as a deterministic logistic ODE perturbed by either environmental fluctuations or by random shifts in the growth rate. We impose $K, \sigma \geq 0$. It is known that this equation has a unique positive definite solution given by

$$X_t = \frac{\exp\left(\sigma W_t + \left(r - \frac{\sigma^2}{2}\right)t\right)}{x_0^{-1} + \frac{r}{K}\int_0^t \exp\left(\sigma W_s + \left(r - \frac{\sigma^2}{2}\right)s\right)ds} \tag{F2}$$

One can recognize how this solution converges to the deterministic case for $\sigma \to 0$. Its probabilistic behavior $P(x,t) := P(x,t|x_0,0)$ is fully described by the Kolmogorov forward equation given by

$$\partial_t P(x,t) = -\partial_x[rx(1 + K^{-1}x)P(x,t)] + \frac{\sigma^2}{2}\partial_{xx}[xP(x,t)]. \tag{F3}$$

Setting the left-hand side to zero and solving the FPE under the normalization requirement given by $\int_0^\infty P(x,t)dx = 1$ and the initial condition $P(x,0) = \delta(x - x_0)$, where $\delta$ is the Dirac delta function, yields the stationary distribution $P(x)$ of (F2). If $\left(r > \frac{\sigma^2}{2}\right)$ holds, letting $t \to \infty$ one obtains the species' stationary distribution $P(x)$ is the Gamma distribution $\Gamma\left(\frac{2r}{\sigma^2} - 1, \frac{\sigma^2}{2}\frac{K}{r}\right)$
given by

$$P(x) := \Gamma\left(\frac{2r}{\sigma^2} - 1, \frac{\sigma^2}{2}\frac{K}{r}\right) = \Gamma\left(\frac{2r}{\sigma^2} - 1\right)^{-1}\left(\frac{\sigma^2}{2}\frac{K}{r}\right)^{\frac{2r}{\sigma^2}-1} x^{\frac{2r}{\sigma^2}} \exp\left(-x\frac{2r}{\sigma^2 K}\right), \tag{F4}$$

Where $\Gamma(a)$ is the (convergent) Gamma function $\Gamma(a) = \int_0^\infty y^{a-1}e^{-y}dy$. If $r < \sigma^2/2$, then the stationary distribution is degenerate, and the species becomes extinct i.e. the process goes to zero with $t \to \infty$ almost surely (thus certainly going below the low threshold $X_t = 1$). Clearly if $r < 0$ then it's trivially seen that the species process $X_t$ converges to 0 asymptotically, but there exist positive growth rate values for which the species still is driven to extinction by the intensity of the environmental fluctuations. The long-run average of the process (F2) is thus

$$E_{t \to \infty}[X_t] = \frac{K}{r}\left(\frac{2r}{\sigma^2} - 1\right)\frac{\sigma^2}{2} \tag{F5}$$

We now are interested in the *survival* probability, which involves setting an absorbing (i.e. "no return") boundary $X_T = 1$, where $T$ is the (random) time of extinction. We therefore restrict the achievable states of the diffusion to $X_t \in [1, \infty)$. With this in mind, let us now calculate the quantity given by

$$S(x) = \int^x \exp\left(-\int^y \frac{2r}{\sigma^2}\left(\frac{1}{z} - \frac{1}{K}\right)dz\right)dy = \int_1^x z^{-\frac{2r}{\sigma^2}} \exp\left(\frac{2r}{K\sigma^2}(z - 1)\right)dz, \tag{F6}$$

since $x > 1$ always. This function is also known as the scale function for the population process driven by (F1), and its corresponding speed function

$$m(x) = \frac{x^{\frac{2r}{\sigma^2}-2}}{\sigma^2} \exp\left(-\frac{2r}{K\sigma^2}(x-1)\right)$$

One can immediately notice that $S(\infty) = \infty$, as the integrals do not converge, implying that the upper boundary of infinity is not achievable. This implies that the probability $p_e(x)$ of the species process $X_t$ with initial state $x$ to reach carrying capacity before being extinct (i.e. hitting 1 "from the right") is given by

$$p_e(x) = \frac{S(x) - S(1^+)}{S(K) - S(1^+)} = \int_1^x z^{-\frac{2r}{\sigma^2}} \exp\left(\frac{2r}{K\sigma^2} z\right) dz \left(\int_1^K y^{-\frac{2r}{\sigma^2}} \exp\left(\frac{2r}{K\sigma^2} y\right) dy\right)^{-1} \quad (F7)$$

Let us now study the expected time to extinction for a general initial population $X_0 = x$. The corresponding expected time to extinction $T(x)$ defined as the first average time $\tau$ at which $X_\tau = 1$:

$$T(x) := g(x) = E \inf_\tau [X_\tau = 1 | X_0 = x], \quad (F8)$$

which is obtained via the boundary value problem

$$-1 = (rx(1 - K^{-1}x)\partial_x g(x) + \sigma^2 x^2 \partial_{xx} g(x), \qquad g(1) = 0, g(\infty) = 1$$

and for the population process starting at carrying capacity (i.e. $x = K$) is given by

$$T(K) = 2\int_1^K y^{-\frac{2r}{\sigma^2}} \exp\left(\frac{2r}{K\sigma^2} y\right) \int_y^\infty \frac{z^{\frac{2r}{\sigma^2}-2}}{\sigma^2} \exp\left(-\frac{2r}{K\sigma^2} z\right) dz dy. \quad (F9)$$

It is a simple exercise to evaluate this nested integral numerically for different parameter values. Figure F1 plots $T(K)$ as a function of $K$, as well as presenting two comparisons between population processes across the threshold $r = \frac{\sigma^2}{2}$, which is the determinant for existence of a non-extinct stationary distribution. The left-hand panel of Figure F1 shows how increasing carrying capacity (here jointly with initial condition, since $x_0 = K$) has substantially decreasing returns to scale with respect to the expected time to extinction $T$.

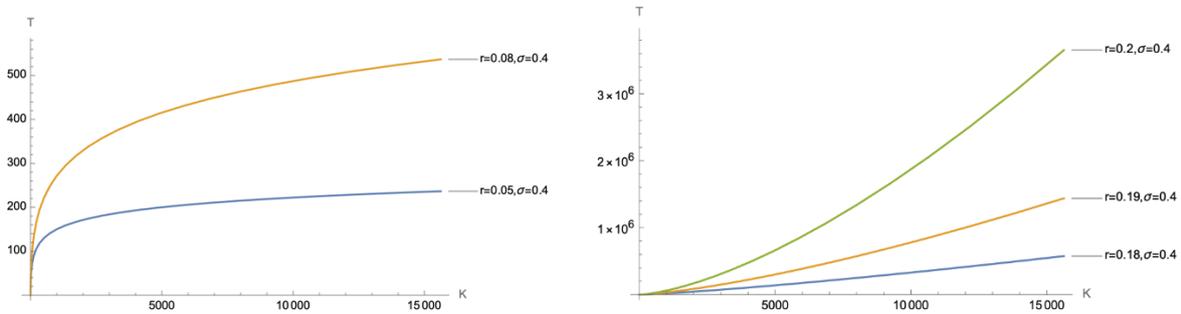

Figure F1: Time to extinction for a population starting at $x_0 = K$ as function of carrying capacity $K$. Left panel: below threshold, right panel: above threshold

This is intuitive: if the ecosystem structure does not allow for a stationary state around the carrying capacity, given by (F5), then increasing initial population level and carrying capacity will not substantially affect the expected time to extinction. On the other hand, the right-hand panel of Figure F1 shows how a system whose population parameters well exceed the threshold exhibits a time to extinction with a convex relation with respect to *K*.

Let us now study the probability of persistence. At any time $t$, the population process $X_t$ obeying (A1) has a transition density $P(x,t)$ satisfying the FPE (F3). The absorbing condition at extinction $X_t = 1$ implies requiring solving (E3) with the added boundary condition

$$P(1,t|x_0,0) = 0,$$

as there must be no probability flow of the process once the boundary is hit, whilst always satisfying the initial condition $P(x,0|x_0,0) = \delta(x - x_0)$.

Solving the stationary FPE with the absorbing boundary condition (F9) yields the following probability distribution $P^{abs}(x)$ at $t \to \infty$ of the species under the extinction threat at $X_t = 1$:

$$P^{abs}(x) = N(\Theta)P^N(x) - \frac{1}{x}\frac{\int_1^\infty z^{r/\sigma^2} \exp\left(\frac{rx}{K\sigma^2}z\right)dz}{\int_1^\infty z^{r/\sigma^2} \exp\left(\frac{r}{K\sigma^2}z\right)dz},$$

where the normalization constant $N(\Theta)$ is a function of the model parameters, and $P^N(x)$ is the unnormalized stationary distribution of the population process without the absorbing boundary at 1.

The "total" survival probability at $t$, $P_S(t|x_0,0)$ implies integrating the transition density $P(x,t)$ over all possible states $x \in [1,\infty)$:

$$P_S(t|x_0,0) = \int_1^\infty P(x,t)dx$$

noticing that

$$P_S(0|x_0,0) = \int_1^\infty \delta(x - x_0), \qquad x_0 > 1,$$

and the *distribution* of the extinction time *T* can be derived from this measure by taking the derivative with respect to time:

$$P_T(t|x_0,0) = -\partial_t P_S(t|x_0,0) = -\int_1^\infty \partial_t P(x,t)dx, \qquad (F10)$$

Using the fact that the KFE must always be satisfied, this means that one can substitute (F3) and obtain

$$P_T(t|x_0,0) = -\int_1^\infty -\partial_x[rx(1 + K^{-1}x)P(x,t)] + \frac{\sigma^2}{2}\partial_{xx}[xP(x,t)]dx$$

$$= (rx(1 - K^{-1}x)P(x,t) - \partial_x[\sigma^2 x^2 P(x,t)])|_{x=1}^{x=\infty}$$

$$= -\sigma^2 x^2 \partial_x P(x,t)|_{x=1} \qquad (F11)$$

as the probability flow *P* is 0 in 1. Since we also have that

$$P_T(0|x_0,0) = \sigma^2 x^2 \delta(x - x_0)|_{x=1} = 0 \tag{F12}$$

then we obtain the probability of "forever persistence" $P_S(\infty|x_0)$ as the integral over all times $t \in [0, \infty)$:

$$\int_0^\infty P_T(t|x_0,0)dt = -\int_0^\infty \partial_t P_S(t|x_0,0) = P_S(t|x_0,0)|_{t=0}^{t=\infty} = 1 - P_S(\infty|x_0). \tag{F13}$$

However, there is no closed form solution for the transition density $P(x,t|x_0,0)$. We can now go further in the analysis by exploiting the fact that the simulations are done starting at carrying capacity, i.e. $x_0 = K$. Intuitively, at $t = 0$ the drift is zero and in $dt$ the increment $dX_0$ is only determined by its diffusive part, which is equal to $y_0 = \sigma x_0 dW_0$. Centering the process on $y \to x - K$, changing time for $Y$ such that $t \to t + dt$ and after a straightforward Lamperti transform, we now focus on the Ito process $Y_t$ given by

$$dY_t = -\rho Y_t dt + \sigma dW_t, \qquad Y_0 = y_0,$$

where $\rho = r/K$, under the absorbing boundary of the extinction threshold $\bar{Y} = K - 1$. This is an Ornstein-Uhlenbeck process, that in the original coordinates is equivalent to a population process initially shocked away from $K$ and reverting towards it at speed $r/K$. The advantage of this approximation is that it allows us to study further the form of (F13), since for the Ornstein-Uhlenbeck the form of the transition density is well known. As per the standard approach, we study the first-passage time distribution (F10) for $Y_t$ into Laplace space:

$$\widetilde{P_T}(\lambda|Y_0) = \int_0^\infty e^{-\lambda t} P_T(t|\bar{Y},0)dt = e^{\frac{\rho(Y_0^2 - (K-1)^2)}{2\sigma^2}} \frac{D_{-\lambda/\rho}\left(-Y_0\sqrt{2\rho/\sigma^2}\right)}{D_{-\lambda/\rho}\left(-(K-1)\sqrt{2\rho/\sigma^2}\right)}$$

where $D_\lambda(z)$ is the parabolic cylinder function, and can be written as a spectral decomposition whose eigenvalues are the zeroes of the denominator $D_{-\lambda/\rho}\left(-(K-1)\sqrt{2\rho/\sigma^2}\right)$ (Ricciardi and Sato, 1988). One can then invert each term to obtain the required passage time distribution $P_T(t|Y_0)$ as an (infinite) weighted sum. Following Giorgini et al. (2020), for large $t$ and "large" $K - Y_0$ (so with $K$ large with respect to $\sigma x_0 dW_0$, which in the original coordinates corresponds to the initial environmental fluctuation $dX_0$ not to reach "too close" to the extinction threshold 1), only the first eigenvalue of the spectrum contributes significantly, and we can write the limit

$$\lim_{K - Y_0 \to \infty} P_T(t|Y_0) = \frac{1}{t_1(K)} \exp\left[-\frac{t}{t_1(K)}\right], \tag{F14}$$

where one has

$$t_1(K) = \frac{K}{r(K-1)} \sqrt{2\pi\sigma^2} \exp\left(\frac{r(K-1)^2}{K\sigma^2}\right).$$

The survival probability for large $K - Y_0$, as shown before, is then given by the time integral of (F14), and is given by

$$P_S(t) \sim \exp\left[-\frac{(K-1)\rho}{\sqrt{2\pi\sigma^2}} \exp\left(-\frac{(K-1)^2\rho}{2}\right)t\right], \tag{F15}$$

which, remembering that $\rho = r/K$ and fixing time to any given amount T, is a Gompertz curve in $K$.

## G. Modifying the standard Gompertz curve

**An example of the difference between the standard and modified Gompertz curves**

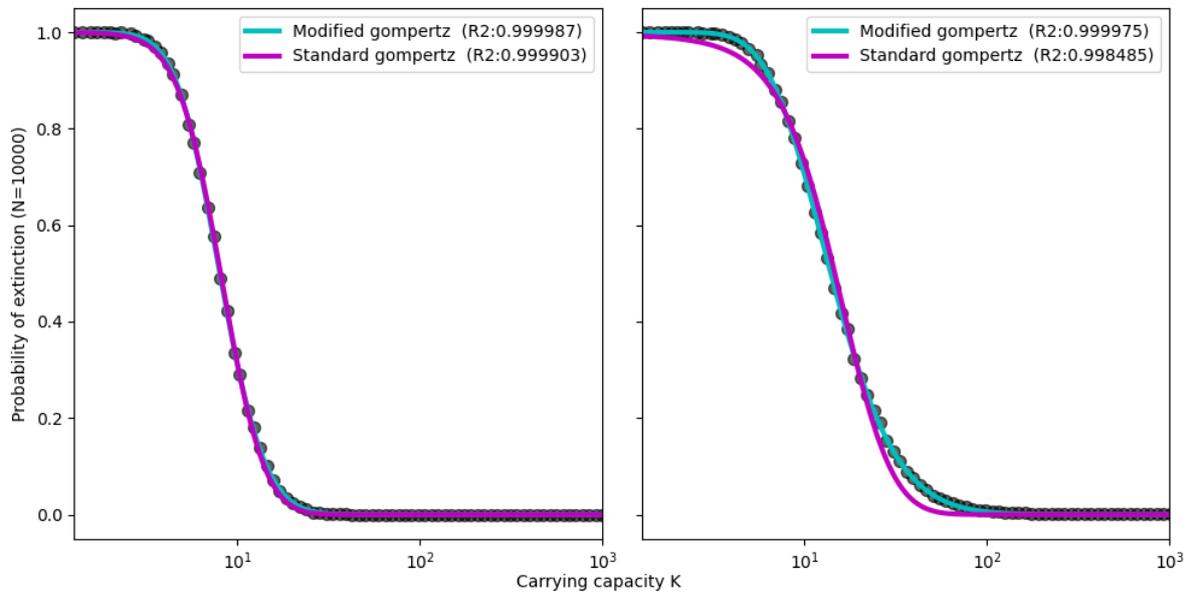

Figure G1. Probability of extinction $P_E$ versus carrying capacity ($K$) with fitted modified and standard Gompertz curves and associated R2 values, for input parameter combinations $r_{max}$ =0.158, $\sigma$=0.11 (left) and $r_{max}$ =0.158, $\sigma$=0.20 (right). The right hand figure illustrates the deviation between the two curves that we observe to increase as $K$ increases – a trend which is persistent across *all* of our models.

**Summary of difference in fit between the standard and modified Gompertz**

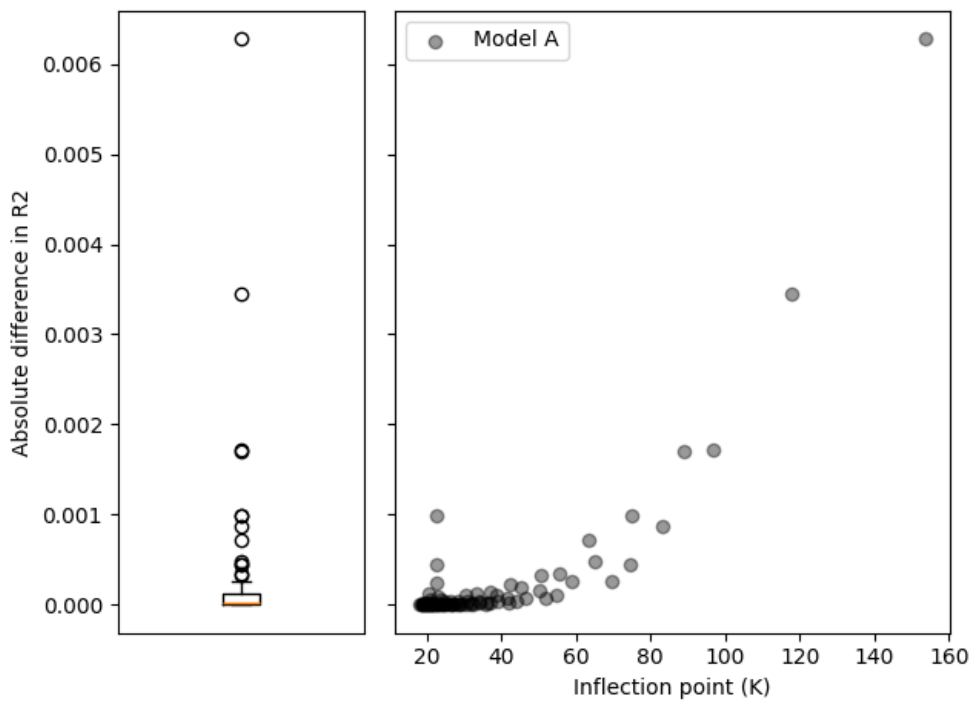

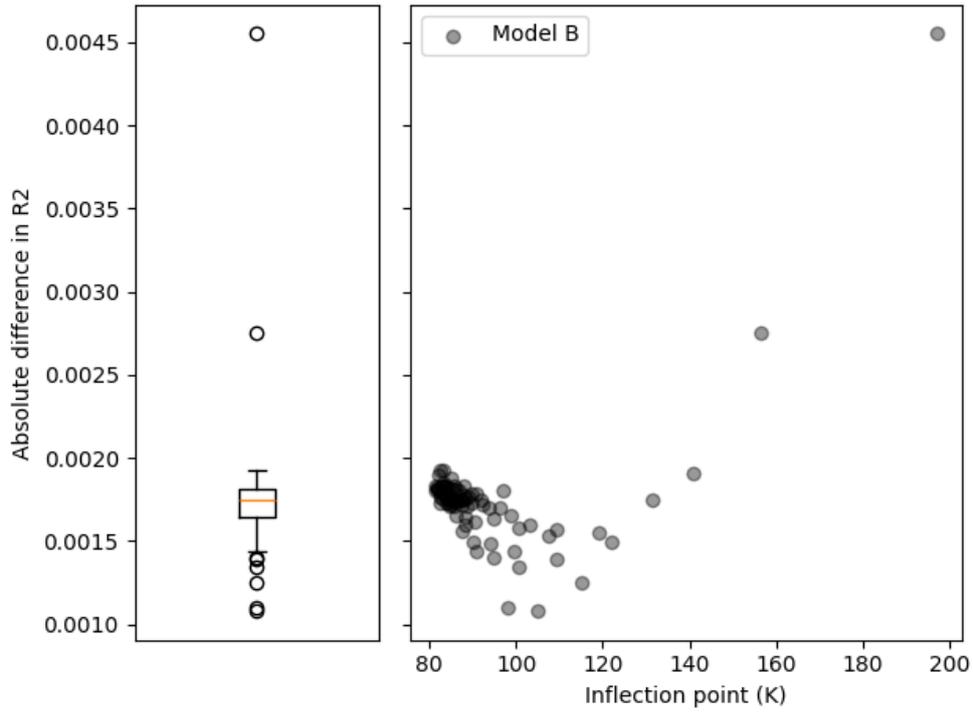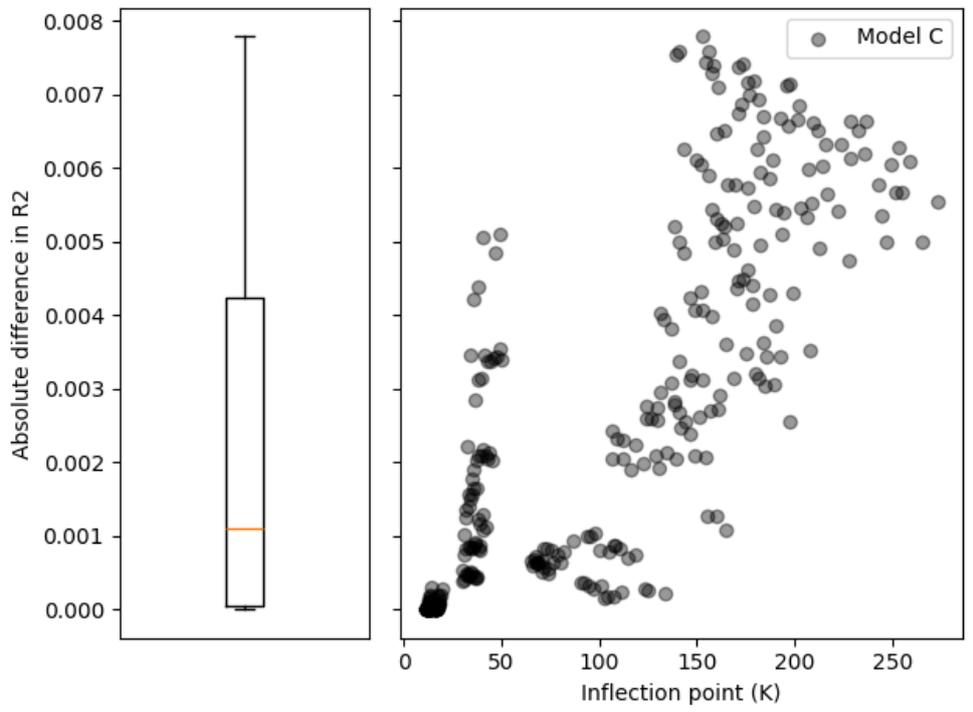

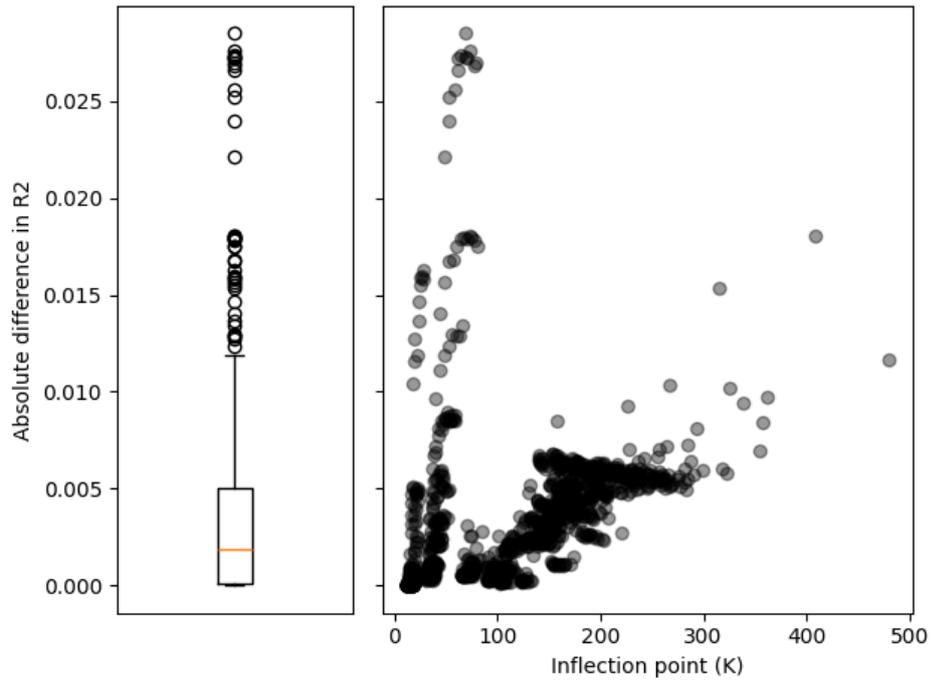

Figures G2-5. For Models A, B, C, and D, the distribution of differences between the R2 values when fitting the modified versus standard Gompertz curves to simulated extinction risk $P_E$ versus carrying capacity $K$ (left), and the same differences versus the inflection point on the modified curve (right). The difference between the two is exclusively positive for all models – the modified curve always has a better fit. Note that, for visual clarity, scaling is not consistent in the y-axis.

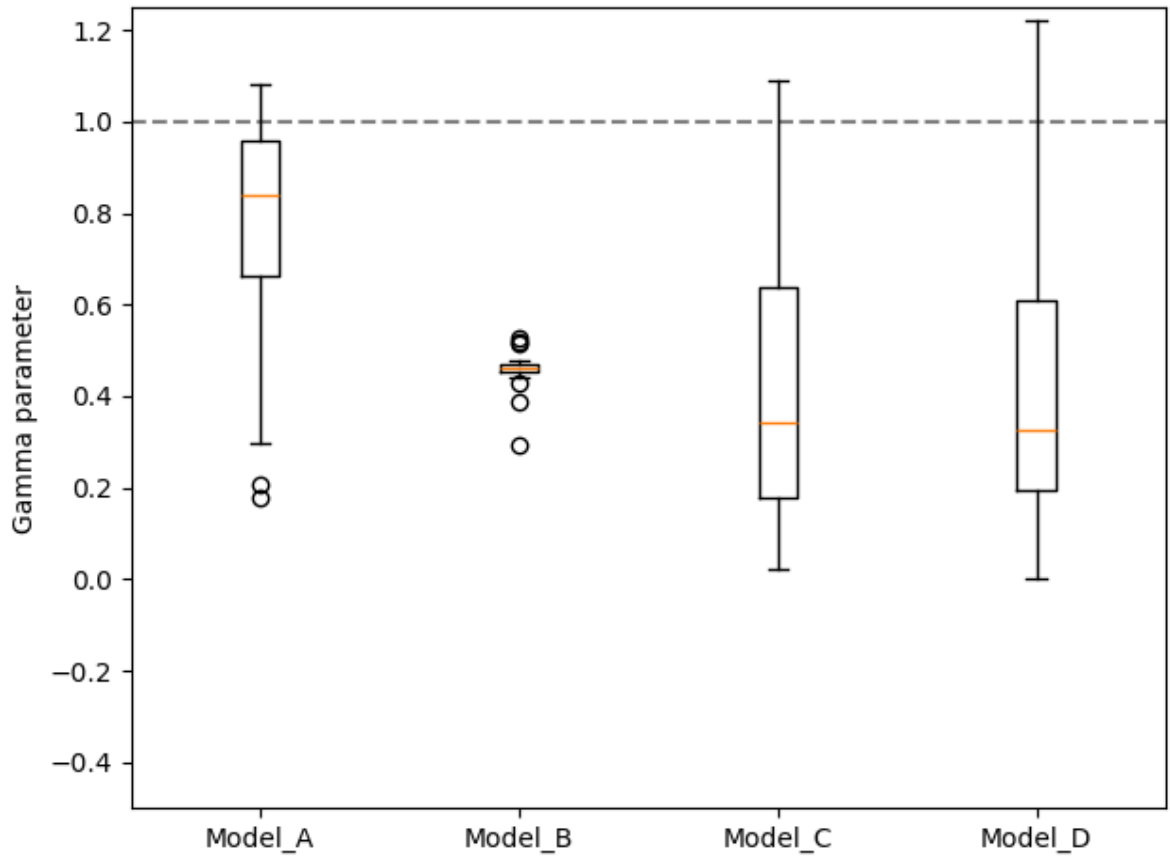

Figure G6. Distribution of the values taken by the shape parameter $\gamma$ in the modified Gompertz curve when fitted to valid results (i.e. for which $\min(P_E) = 0$) across simulation models A – D. Boxes represent the 25[th] to 75[th] percentiles, whiskers represent 1.5x the interquartile range, values outside that range are represented by dots. Horizontal lines represent the 50[th] percentile (median).

## H. Inflection point in the extinction probability axis

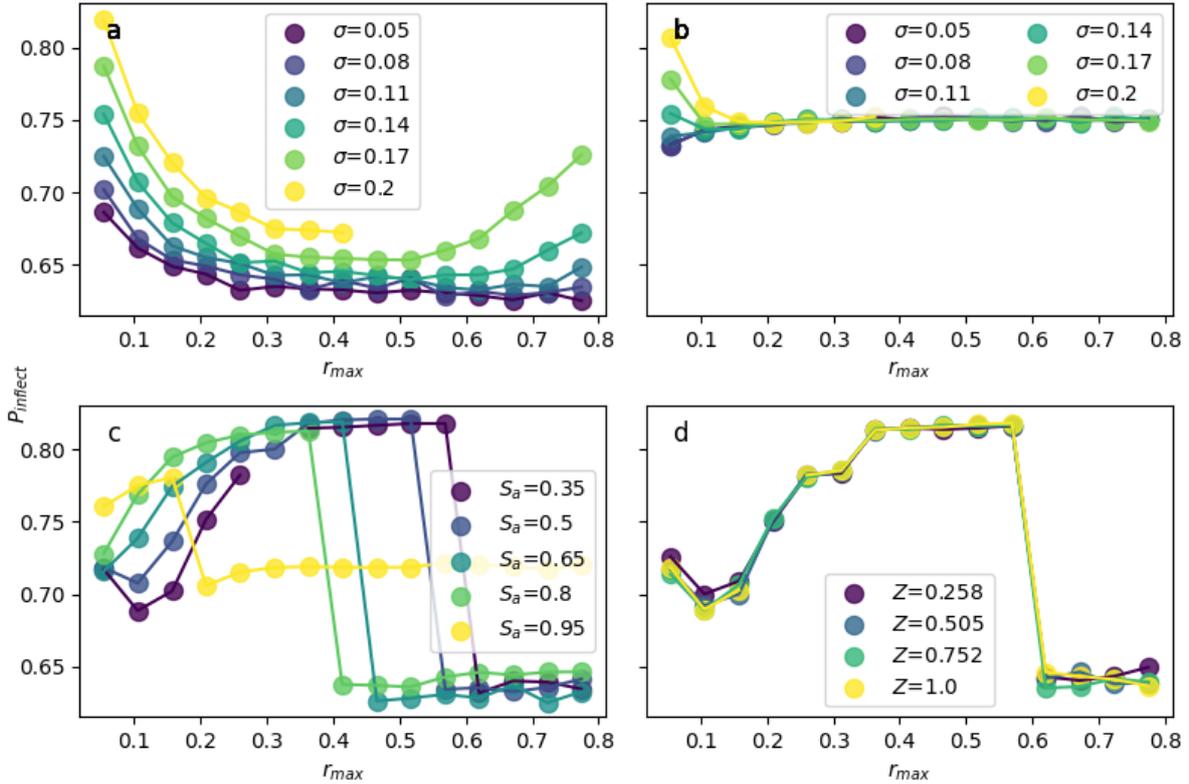

Figure H1. The $P_E$ value at the point of inflection ($P_{\text{inflect}}$) on the fitted extended Gompertz curve against $r_{\max}$, when varying $\sigma$ in (a) Model A and (b) Model B. Then (c) fixing σ = 0.08 but varying $S_a$ a in Model C. Finally, (d) setting $S_a$ = 0.35 and σ = 0.08 but varying $Z$ in Model D.

## I. Compatibility with the Red List criteria

**Rationale underlying the calculation of $K_{10}$, the carrying capacity at which the expected probability of extinction in 100 years $P_E$ is 0.1**

In addition to using our simulation results to describe relationships between $P_E$ and $K$, we also wished to assess the effects of values of input parameters on a single summary statistic chosen to represent variation among modelled species' populations in their susceptibility to extinction during a period of 100 years. For this purpose, we chose to estimate, based upon our fitted Gompertz functions, the $K$ value for which the probability of extinction in 100 years was 0.1 (10%). We call this $K_{10}$. We chose this particular statistic because the same probability is also used in IUCN Red List classification as the only criterion (Criterion E) which explicitly refers to the probability of extinction in a defined time period. To classify a species as Vulnerable, which is the least endangered of the three categories of species considered to be at risk of global extinction which are not already Extinct or Extinct in the Wild, Criterion E requires that the modelled probability of extinction in 100 years is 10% or greater (IUCN 2001). An extinction probability of 5% or 10% in 100 years was initially suggested as an acceptable level of risk by Shaffer (1981). Criterion E definitions for the other two categories (Critically Endangered and Endangered) involve higher probabilities of extinction over a comparable period than does Vulnerable, but the definitions are more complicated and therefore more difficult to link with our simulation results.

We therefore think that $K_{10}$ is a straightforward and appropriate statistic that aligns well with the IUCN Red List criterion referring to extinction risk which separates all globally threated species from other, less threatened species (Near Threatened and Least Concern), given that a species' population is potentially at risk because of consequences of having small population size due to low carrying capacity, rather than from having a population undergoing a sustained decline. We calculated $K_{10}$ from the three fitted parameters of the Gompertz function ($a$, $b$ and $\gamma$) as

$$K_{10} = \left(\frac{\ln(-\ln(0.1)) - a}{b}\right)^{\frac{1}{\gamma}} \quad (I1)$$

We examined the relationship between our simulation parameters and $K_{10}$. Figures I1 and I2 shows the relationship between $K_{10}$ and input parameters. For Models A and B, $K_{10}$ decreases with increasing $r_{\max}$, but is relatively invariant across population parameter combinations for $r_{\max}$ values >0.4. The $K_{10}$ value of the plateau is approximately 200 individuals greater for Model B than A, and in both model types increases with environmental stochasticity ($\sigma$). In Model C, varying $S_a$ against $K_{10}$ has a hump-shaped relationship to $r_{\max}$ (Figure G1a) with lower values of adult survival $S_a$ shifting the position of the hump toward lower $r_{\max}$ values. In Model D, $K_{10}$ increases slightly with decreasing values of $Z$, reflecting greater temporal autocorrelation in the environmental stochasticity faced by the population, with the strongest impacts at $r_{\max}$ = 0.4.

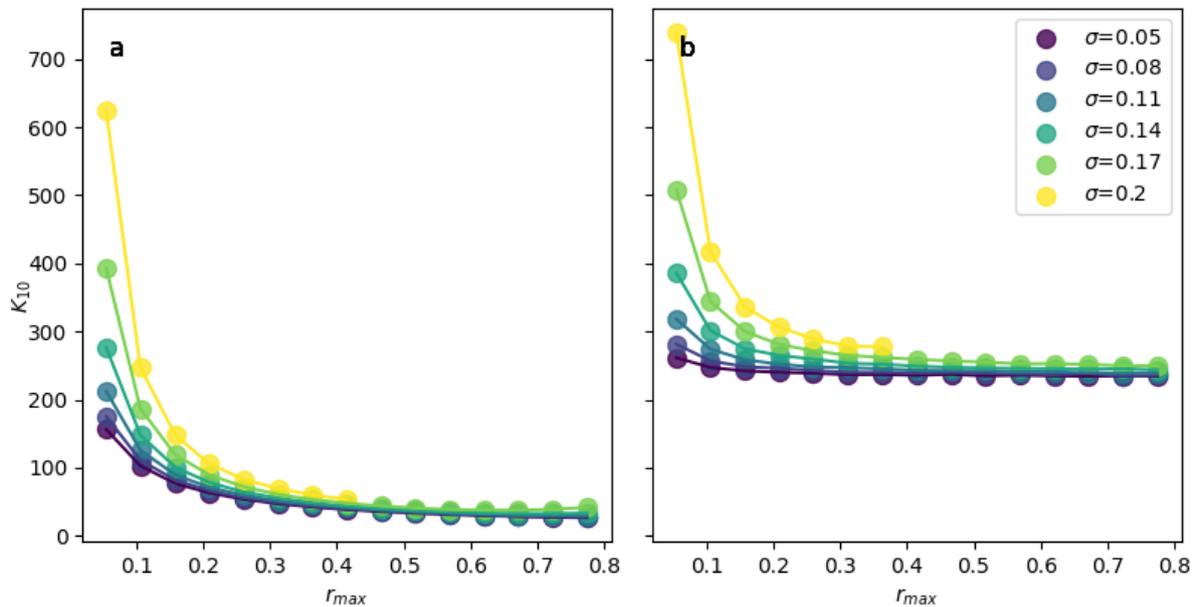

Figure I1. The relationship between $K_{10}$ – the $K$ value at which $P_E$ = 0.1 - and maximum growth rate $r_{\max}$ for various values of environmental stochasticity $\sigma$ for (a) Model A and (b) Model B.

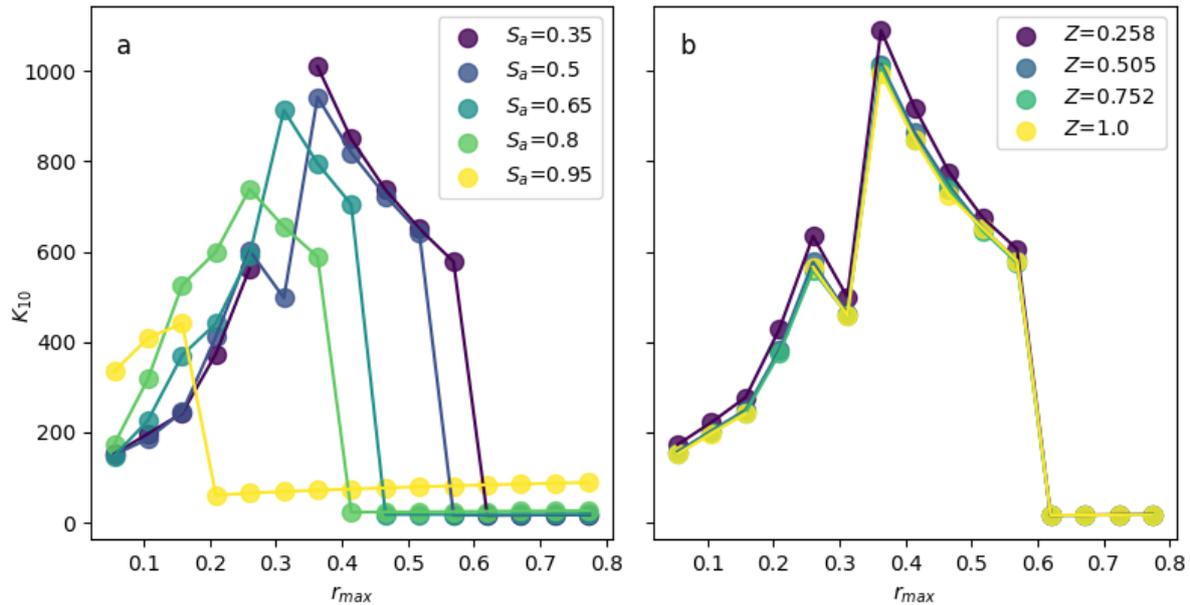

Figure I2. The relationship between $K_{10}$ and $r_{\max}$ for varying values of $S_a$ in (a) Model C, and the relationship between $K_{10}$ and $r_{\max}$ for various values of $Z$ and a constant $S_a$ = 0.35 in (b) Model D.

**Degree to which the modelled parameter K10 is representative of extinction risk more widely**

We calculated the value of a single parameter $K_{10}$ from values of the parameters *a*, *b* and *Y* of the modified Gompertz function fitted to each of our simulations. $K_{10}$ is the expected *K* value at which the modelled probability of extinction within 100 years is 0.1 (10%). Our choice as an extinction risk metric of the *K* at which extinction probability is 10% aligns with IUCN's Red List Criterion E (see main text), but chosen extinction risk level is arbitrary. We therefore wished to assess the degree to which variation in $K_{10}$ among our simulations was representative of extinction risk for the same simulation more widely. To do this, we calculated values of $K_{10}, K_{20}, K_{30}, K_{40}, K_{50}, K_{60}, K_{70}, K_{80}$ and $K_{90}$ for each of the simulations obtained for a given model (Models A to D) and then plotted each of $K_{20}, K_{30}$, etc against $K_{10}$. These plots showed that the relationship of $K_i$ to $K_{10}$ was curvilinear and concave for all four models. In each case, it was possible to linearise the relationship by transforming $K_{10}$ to $K_{10}{}^c$, where *c* is a constant >1. We determined the optimal value of *c* by using a bisection search method (Kalbfleisch, 2011) to find the value of *c* at which the Pearson correlation coefficient *r* between $K_i$ and $K_{10}{}^c$ was maximised. Table J1 shows the results of this analysis and Figure X.1 shows the square of the Pearson correlation coefficient *r* ($r^2$) plotted against *i* (the percentages for $K_i$; *i* = 20, 30,... 90). For all four models, there was a tendency for the correlation to become weaker as the difference between *i* and 10 in $K_i$ became larger (Figure I2). However, it was possible to predict $K_i$ from $K_{10}$ reasonably well for all four models, but especially for Models A and B (Figure I2).

Table I1. Optimal values of the shape parameter *c* and the Pearson correlation coefficient *r* between $K_i$ and $K_{10}{}^c$ for different values of the extinction risk *i*, expressed as a percentage.

| | Model A | | Model B | | Model C | | Model D | |
|---|---|---|---|---|---|---|---|---|
| *i* (%) | c | r | c | r | c | r | c | r |
| 20 | 1.31 | 0.9999 | 1.67 | 0.9994 | 1.01 | 0.9959 | 1.04 | 0.9951 |

| 30 | 1.55 | 0.9997 | 2.14 | 0.9984 | 1.03 | 0.9888 | 1.08 | 0.9864 |
| 40 | 1.77 | 0.9996 | 2.55 | 0.9976 | 1.05 | 0.9807 | 1.13 | 0.9764 |
| 50 | 1.99 | 0.9995 | 2.94 | 0.9967 | 1.08 | 0.9719 | 1.18 | 0.9652 |
| 60 | 2.23 | 0.9993 | 3.32 | 0.9960 | 1.12 | 0.9623 | 1.24 | 0.9527 |
| 70 | 2.50 | 0.9990 | 3.73 | 0.9952 | 1.16 | 0.9515 | 1.31 | 0.9376 |
| 80 | 2.85 | 0.9986 | 4.23 | 0.9945 | 1.23 | 0.9384 | 1.40 | 0.9174 |
| 90 | 3.39 | 0.9981 | 4.92 | 0.9937 | 1.33 | 0.9197 | 1.53 | 0.8802 |

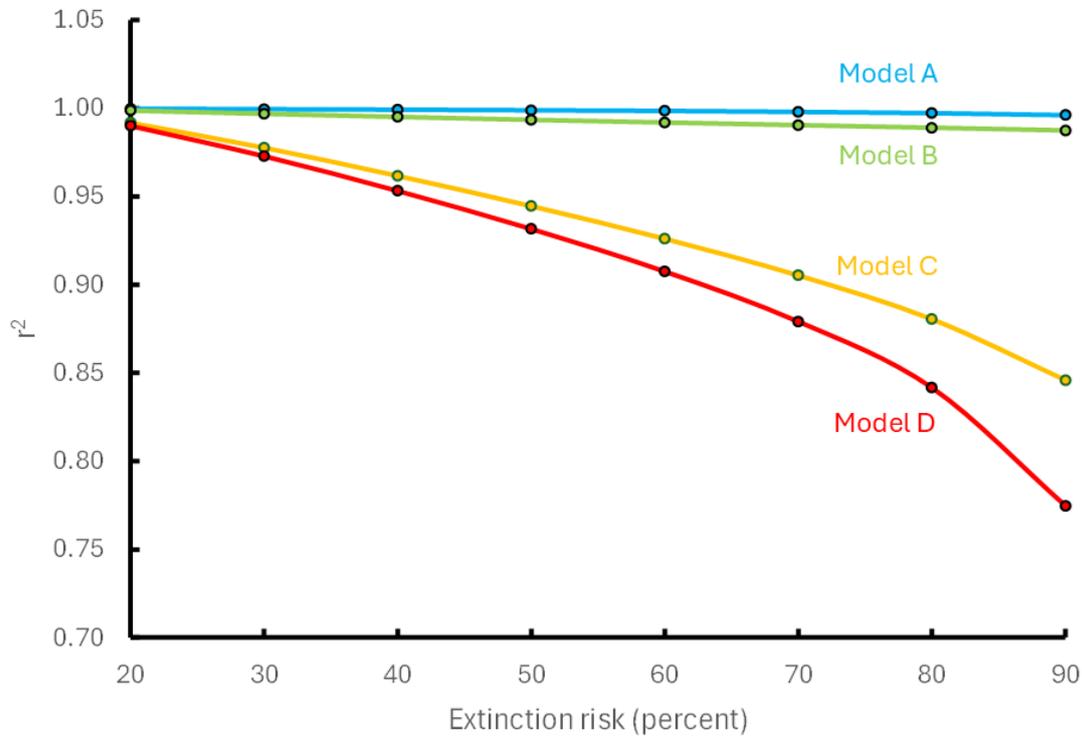

Figure I3. The square ($r^2$) of the Pearson correlation coefficient r between $K_i$ and $K_{10}^c$ in relation to the value *i* of extinction risk in $K_i$, expressed as a percentage.

Whilst the choice of 10% by the Red List process is arbitrary, variation in $K_{10}$ values among our simulations were quite well correlated with other extinction threshold $K$ values, especially for Models A and B. We thus conclude that the expected value of $K_{10}$ is a reasonable proxy for variation in extinction risk among species and populations.

## J. Predicting features of the curve

**Predicting $K_{10}$ values from input parameters using linear models**

We used the scikit-learn Python package (Pedregosa 2011) to build linear models for predicting $K_{10}$, $K_{50}$, and $K_{90}$ values for all four models from simulation input parameters ($r_{\max}, \sigma, S_a,$ and $Z$). Predicted versus observed values and the associated $r^2$ values for each of these are shown in Figures J1-3 below, and the model parameters themselves can be found in the supplementary data repository.

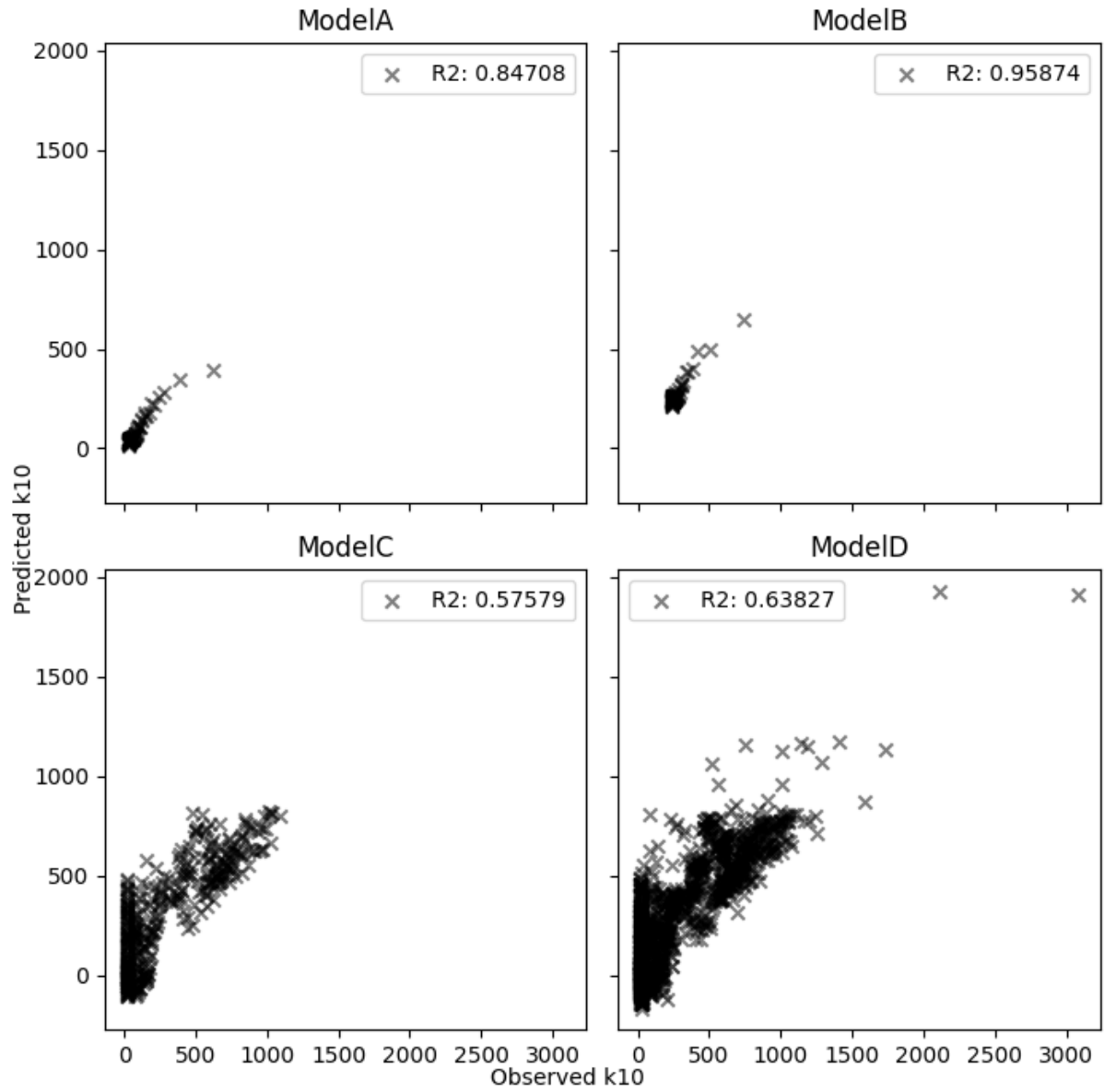

Figure J1. $K_{10}$ values as predicted by our linear model versus observed $K_{10}$ value from our simulated results.

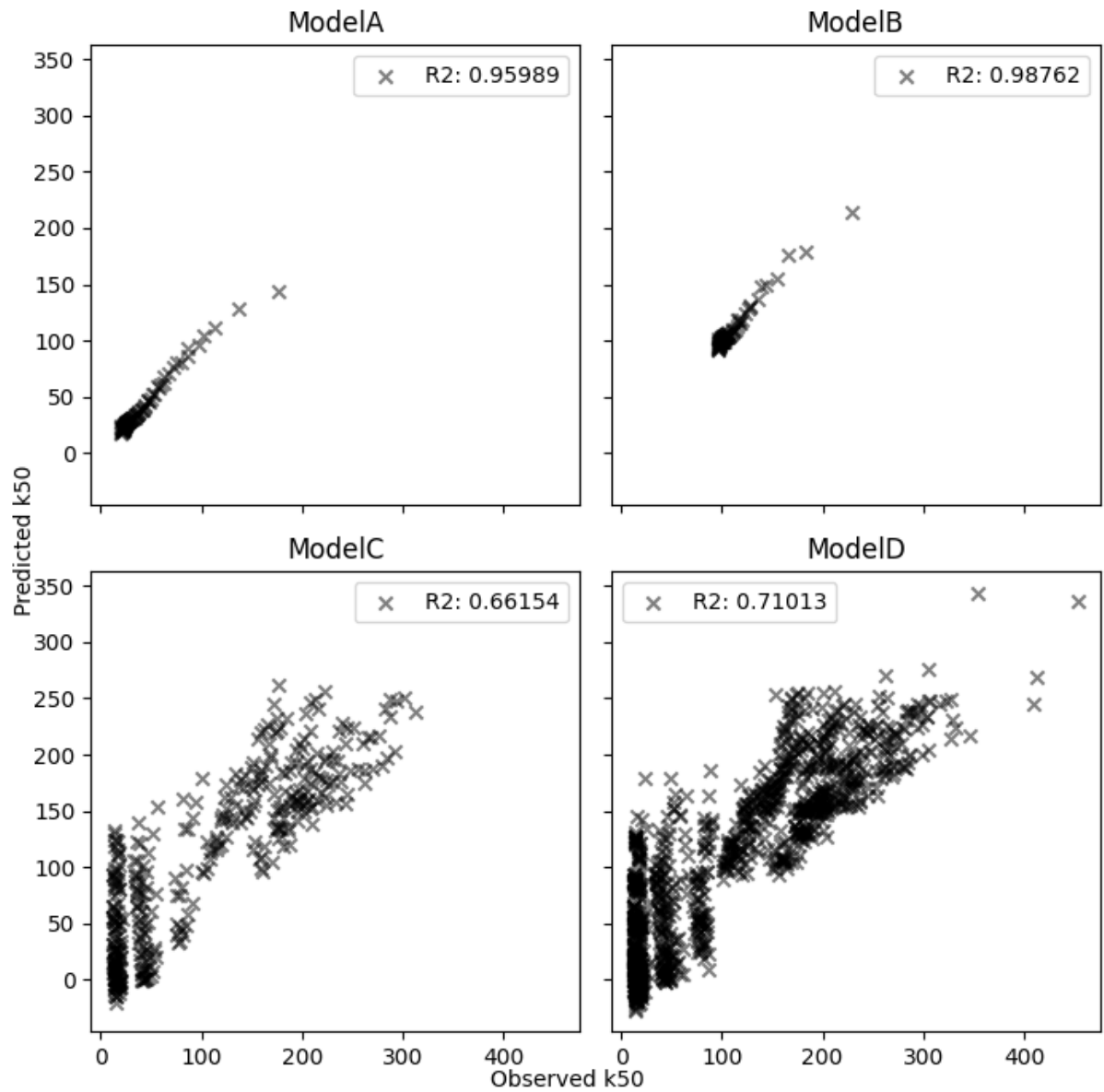

Figure J2. $K_{50}$ values as predicted by our linear model versus observed $K_{50}$ value from our simulated results.

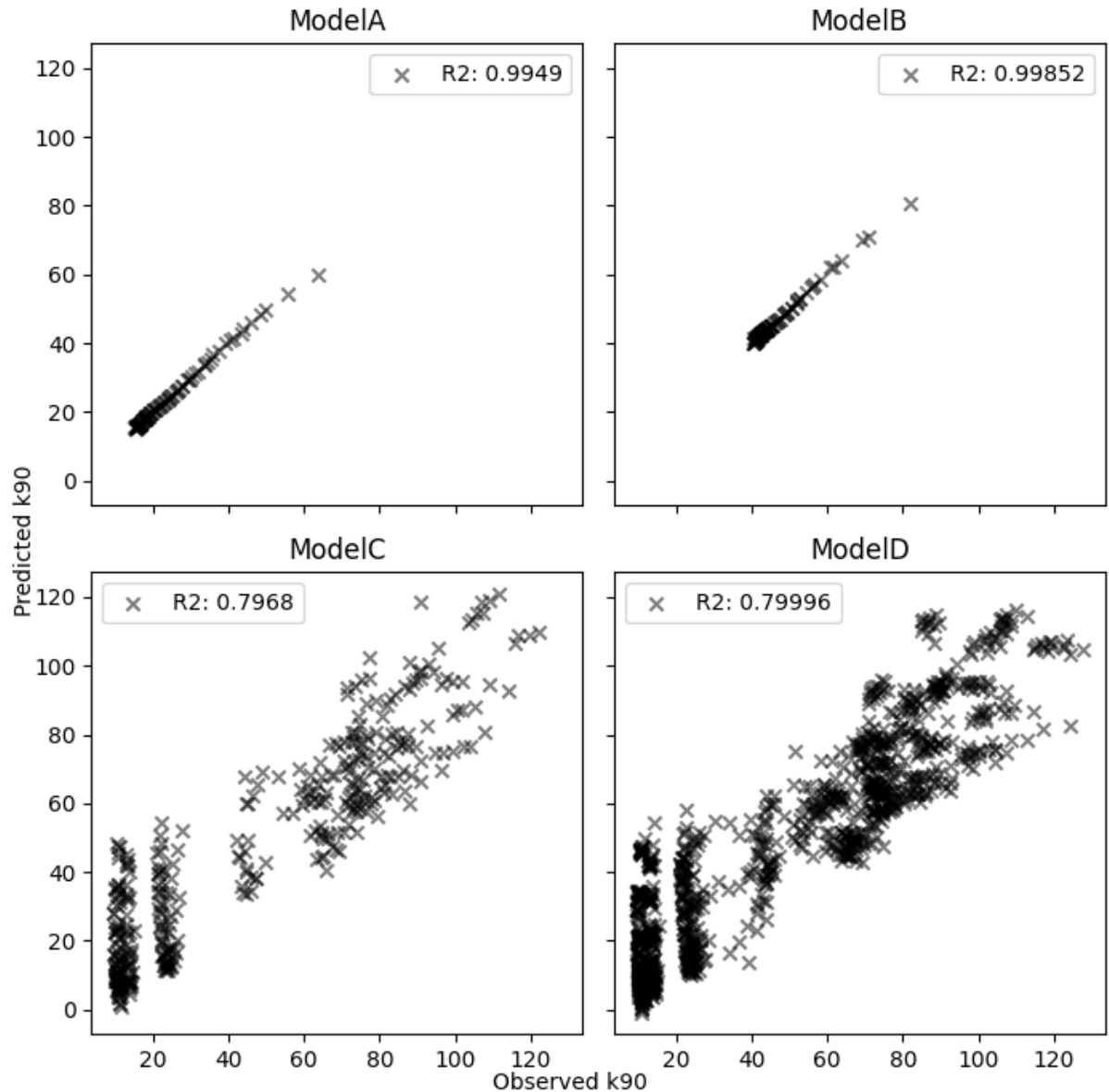

Figure J3. $K_{90}$ values as predicted by our linear model versus observed $K_{90}$ value from our simulated results.

**Predicting the standard Gompertz parameters**

Ideally, we would construct simple linear models to predict features of the modified Gompertz from simulation input parameters. Doing so unfortunately suffered from an under-identification issue. So we instead predicted the standard, unmodified Gompertz curves, which we were able to predict directly from life history parameters with high accuracy for models A and B ($r^2$ between 0.91 and 0.99) and moderate accuracy for models C and D ($r^2$ between 0.73 and 0.84). As previously discussed, the standard Gompertz systematically underestimates extinction risk at high carrying capacities, but we suggest that even an approximation of the true relationship from life history parameters is useful. Further, we found that we could predict the extinction threshold carrying capacities discussed in the previous section with reasonable accuracy. We modelled $K_{10}$, along with $K_{50}$ and $K_{90}$ (50% and 90% probabilities of extinction respectively), the former of which is related to IUCN's Critically endangered category, albeit for different timescales (IUCN 2001). We were able to predict the values of $K_{10}$, $K_{50}$, and $K_{90}$ with variable success across our simulation models. For Models A and B, we achieved a predicted versus observed r² between 0.85 – 0.99 when predicting the $K$ values, whereas for C and D we achieved r² values between 0.57 and 0.79 (see Supplementary Appendix J for details).

We used the scikit-learn python package (Pedregosa 2011) to build linear models for predicting the two parameters $a$ and $b$ in the standard Gompertz equation:

$$1 - P_E = \exp(-\exp(a + bK)), \tag{J1}$$

equivalent to the modified Gompertz (Equation 1 in the main text) for which the shaping parameter $\gamma$ is set to 1. We predicted the $a$ and $b$ values for the fitted standard Gompertz for all four simulation models from their input parameters ($r_{\max}, \sigma, S_a,$ and $Z$). Associated r² values and linear model coefficients can be found in the supplementary data repository. Though we were unable to reliably predict the modified Gompertz from life history parameters in this instance, it may well be possible to do so, and is an avenue for future work.